\documentclass[aps,showpacs,floatfix,superscriptaddress,11pt,nofootinbib]{revtex4-1}
\usepackage{graphicx}
\usepackage{amssymb}
\usepackage{graphics, color}
\usepackage{overpic}
\usepackage{float}
\usepackage{hyperref}
\usepackage[english]{babel}
\usepackage[usenames,dvipsnames]{xcolor}
\usepackage{amsmath,amsfonts,amssymb,verbatim,float,mathtools}

\pdfoutput=1

\newcommand{\axa}{A_x^{(0)}}
\newcommand{\axb}{A_x^{(1)}}
\newcommand{\tBx}{\tilde B_x}
\newcommand{\tgxt}{\tilde g_{xt}}
 \newcommand{\gxt}{g_{xt}}
  \newcommand{\gxta}{{g_{xt}^{(0)}}}
  \newcommand{\gxtb}{g_{xt}^{(1)}} 
  \newcommand{\gtt}{g_{tt}}
  \newcommand{\guu}{g_{uu}}
\newcommand{\gxx}{g_{xx}}
  \newcommand{\OO}{{\cal O}}
  \newcommand{\Sax}{S_{\text{axion}}}
  \newcommand{\Semd}{S_{\text{EMd}}}
  \newcommand{\sigmadc}{\text{Re}(\sigma_{\rm DC})}
  \newcommand{\be}{\begin{equation}}
  \newcommand{\ee}{\end{equation}}
  \newcommand{\bea}{\begin{eqnarray}}
\newcommand{\eea}{\end{eqnarray}}

\def\ba{\begin{array}}
\def\ea{\end{array}}

\newcommand*\pFq[2]{\;{}_{#1}F_{#2}}

\newcommand{\ch}{\mathrm{\cosh}}
\newcommand{\sh}{\mathrm{\sinh}}
\newcommand{\rh}{r_H}

\definecolor{Gray}{gray}{0.4}

\hypersetup{
     colorlinks   = true,
     citecolor    = blue,
    linkcolor=blue,
    urlcolor=blue
}

\begin{document}
\title{Drude weight and Mazur-Suzuki bounds in holography}

\author{Antonio M. Garc\'\i a-Garc\'\i a}
\author{Aurelio Romero-Berm\'udez}
\affiliation{TCM Group, Cavendish Laboratory, University of Cambridge, JJ Thomson Avenue, Cambridge, CB3 0HE, UK}

\begin{abstract}
We investigate the Drude weight and the related Mazur-Suzuki (MS) bound in a broad variety of strongly coupled field theories with a gravity dual at finite temperature and chemical potential. We revisit the derivation of the recently proposed universal expression for the Drude weight for Einstein-Maxwell-dilaton (EMd) theories and extend it to the case of theories with multiple massless gauge fields. We show that the MS bound, which in the context of condensed matter provides information on the integrability of the theory, is saturated in these holographic theories including R-charged backgrounds.
We then explore the limits of this universality by studying EMd theories with $U(1)$ spontaneous symmetry breaking and gravity duals of non-relativistic field theories including an asymptotically Lifshitz EMd model with two massless gauge fields and the Einstein-Proca model. In all these cases, the Drude weight, computed analytically, deviates from the universal result and the MS bound is not saturated. In general it is not possible to deduce the low temperature dependence of the Drude weight by simple dimensional analysis. Finally we study the effect of a weak breaking of translational symmetry by coupling the EMd action, with and without $U(1)$ spontaneous symmetry breaking, to an axion field. We show the coherent part of the conductivity in this limit is simply the product of the MS bound and the scattering time obtained from the leading quasinormal mode. For asymptotically $AdS$ theories it seems that the MS bound sets a lower bound on the DC conductivity for a given scattering time.  
\end{abstract}
\date{\today}

\maketitle
\section{Introduction}
Momentum is conserved in the absence of interactions, impurities and lattice defects. Transport is ballistic and the material is a perfect conductor with an electrical conductivity that diverges in the limit of vanishing frequencies. This is in principle a highly idealized situation as even for good metals there are different mechanisms of momentum relaxation, from impurities and electron-electron interactions to Umklapp scattering, that render the transport diffusive. 
It is therefore plausible to expect that quantum ballistic motion, especially at finite temperature and in the presence of a lattice, cannot occur in a strongly interacting system. However, this is far from being true \cite{zotos1997,fujimoto1998,fujimoto2003,karrasch2012,sirker2011}. 
Paradigmatic examples of one dimensional systems with this property in a broad range of parameters are the spin $1/2$-XXZ chain \cite{zotos1999,fujimoto2003} or the repulsive Hubbard model \cite{fujimoto1998}. Ballistic transport is usually characterized by  the strength of the delta function in the conductivity for vanishing frequencies, the so called Drude weight $K$.

Explicit analytical expressions in some interacting one-dimensional systems \cite{zotos1999} are available 
by expressing the Drude weight \cite{kohn1964} as a function of the flux dependence of the spectrum, which is obtained by Bethe ansatz.
Monte Carlo and DMRG techniques \cite{karrasch2012}, together with a finite size scaling analysis, have also been heavily used to determine the conditions for a finite Drude weight to occur and its explicit temperature dependence. Despite these advances there are still conflicting results in the literature, see \cite{sirker2011,karrasch2012} about the exact range of parameters in which transport is ballistic. This is not surprising as the extrapolation of the numerical results to the thermodynamic limit is especially challenging in the case of the direct current conductivity and analytical approaches contain reasonable but uncontrollable approximations. 

However, the very existence of a finite Drude weight is in many cases guaranteed by the Mazur-Suzuki (MS) bounds \cite{mazur1969,suzuki1971}. These bounds relate \cite{zotos1997} the Drude weight to a weighted positive definite sum of correlation functions between the electrical current and the conserved quantities of the system.

A sufficient condition for a finite Drude weight is thus the existence of some overlap between the current and a single conserved quantity. More recently, Mazur-Suzuki bounds have been generalized \cite{prosen2011,ilievski2013} to quasi-local
conservation laws and systems with open boundary conditions. The new derivation of the bounds \cite{ilievski2013} is heavily based on causality constraints as given by Lieb-Robinson bounds \cite{lieb1972}. A finite bound is also deeply related to the non-ergodicity of the operator in question, in this case the current though it can be generalized no any other bound observable.  More explicitly in Refs. \cite{castella1995,zotos1997} it was conjectured that a finite Drude weight implies non-ergodicity of the dynamics, and consequently, some form of quantum integrability of the model. In case that the Mazur-Suzuki bound is saturated it was recently  proposed \cite{mierzejewski2014} that the thermodynamic properties of the system are well described by the generalized Gibbs ensemble. 

So far, severe technical limitations, both analytical and numerical, have prevented a systematic study of Mazur-Suzuki bounds and Drude weights in higher dimensional systems. An important exception are strongly coupled theories with a gravity dual \cite{maldacena1999,hartnoll2009a,sachdev2011}, where Drude weights have been computed analytically in many situations \cite{jain2010,chakrabarti2011,gubser2010,Son2006,Davison2015a,Kiritsis2015,Hartnoll2007,Hartnoll2007b,Jottar2010,Hoyos2010,
Dewolfe2011,Dewolfe2012}.
For instance the Drude weight for Einstein-Maxwell theories with a single massless gauge field was discussed in Refs. \cite{Hartnoll2007b,Hartnoll2007}. The Drude weight in some R-charged backgrounds was worked out in \cite{Dewolfe2011,Dewolfe2012} and in \cite{Hoyos2010} for probe D-branes in a Lifshitz space time. 
The calculation of the Drude weight in more general Einstein-Maxwell theories and the proposal of universality was first made in \cite{jain2010,chakrabarti2011} and then revisited in \cite{Davison2015a}. For holographic superconductors, the Drude weight was computed numerically in \cite{Hartnoll2008}. In the context of holographic theories, it is rather unclear whether a finite Drude weight at finite temperature is related to integrability. However, we note that there are recent claims \cite{Giataganas2014,Klemm2015} that asymptotically $AdS$ Einstein-Maxwell theories are classically integrable.

Here, we extend these studies to the calculation of MS bounds and also Drude weights in a broader ensemble of holographic theories at finite temperature and chemical potential. More specifically, in the first part of the paper we investigate Einstein-Maxwell-dilaton gravity theories with and without $U(1)$ symmetry breaking, R-charged backgrounds, multiple massless gauge fields and gravity theories with massive gauge fields and EMd theories with a non-$AdS$ boundary for which the dual field theories are non-relativistic. These  are known \cite{charmousis2010} to be a fertile ground for phenomenological approaches to condensed matter systems. Indeed we have found a rich phenomenology. For the models where the Drude weight $K$ is given by the universal expression \cite{jain2009,Davison2015a}, that depends only on thermodynamic quantities, the MS bound is saturated. The temperature dependence of $K$ is not, in general, given by simple dimensional analysis. In the case of $U(1)$ scalar condensation the Drude weight is larger than the universal prediction and the MS bound is finite but it is not saturated.
For the non-relativistic field theories we have investigated the MS bound vanishes and the Drude weight is different from the universal prediction. In the second part of the paper we study the DC conductivity once momentum conservation is weakly broken in EMd-axions models with and without $U(1)$ symmetry breaking. We show that the coherent part of the DC conductivity is controlled by the MS bound and the scattering time, which is obtained independently by an explicit calculation of the leading quasinormal mode. At least for asymptotically $AdS$ theories it seems that the MS bound sets a lower non-trivial bound on the DC conductivity for a given scattering time.

The organization of the paper is as follows: next we review previous holography literature on the Drude weight, and revisit the analytical calculation of the universal Drude weight \cite{Davison2015a,jain2010,chakrabarti2011}. In section three we extend this result by proposing a universal Drude weight for theories for multiple massless gauge fields. 
In section four we introduce Mazur-Suzuki bounds and detail how to compute them in some EMd holographic theories. 
The calculation of the Drude weight and the MS bounds in EMd theories with $U(1)$ symmetry breaking and with a non-relativistic dual field theory, where universality does not apply, is carried out in sections five and six.  
Finally, in section seven we study EMd-axion models, with and without  $U(1)$ symmetry breaking, in the limit of weak breaking of translational invariance, where we show that the DC conductivity is controlled by the leading quasinormal mode and the MS bound for the Drude weight. 

\section{Universality of the Drude weight revisited}\label{sec:Universiality}
Although the holography literature has focused mostly on the calculation of the finite part of the DC conductivity, the infinite part, characterized by the Drude weight, has also already received some attention \cite{jain2010,chakrabarti2011,gubser2010,Son2006,Davison2015a,Kiritsis2015,Hartnoll2007,Hartnoll2007b,Jottar2010,Hoyos2010,
Dewolfe2011,Dewolfe2012}. 

Interestingly, the Drude weight corresponding to a single massless gauge field in a Einstein-Maxwell-dilaton theory has been found to be universal and given only by thermodynamic quantities \cite{jain2010,chakrabarti2011,Davison2015a}.
In these papers the focus was on the study of universal aspects of the finite, or regular, part of the DC conductivity, usually referred to as $\sigma_Q$, rather than the Drude weight, though the latter was also computed explicitly. 
We start our analysis by revisiting the derivation of this universal DC conductivity. We adapt it to the analytical calculation of the Drude weight as this is the starting point for the generalization of these results in the following sections. We will follow closely the approach of \cite{Davison2015a} though with some modifications so that the calculation of the Drude weight is more direct and easier to generalize beyond universality. 
The slightly different method of Jain and co-workers \cite{jain2010,chakrabarti2011}, proposed earlier, leads to exactly the same results.

The full DC conductivity is given by the current-current Green's function, \cite{Bradlyn2012},
\be\label{sigma:def}
\sigma_{\rm DC} =-\text{Re}  \lim_{\omega \to 0,q\to0} \frac{\, G^R_{J^x J^x}(\omega,q)-G^R_{J^x J^x}(0,q)}{i(\omega+i\epsilon)}\ ,
\ee
that physically represents the linear response of the system to an external small field perturbation, $a_x$. 
We note that this form of the Kubo formula ensures, that the limit ${\omega \to 0}$, $\epsilon\to0^+$ captures the full paramagnetic response. More specifically, assuming $\lim_{q\to0} G^R_{J^x J^x}(\omega,q)-G^R_{J^x J^x}(0,q)\approx K - i\omega \sigma_{\rm Q}+{\cal O}(\omega^2)$, eq. (\ref{sigma:def}) leads, for $\epsilon\to0^+$, to
\begin{equation}\label{Dirac:form}
\begin{split}
\sigma_{\rm DC}=&\sigma_{\rm Q}-{\rm Re}{1\over i}{K\over \omega+i\epsilon}=\sigma_{\rm Q}-{\rm Re}\left[{\cal P}\left({1\over\omega}\right)(-i)K-\pi K \delta(\omega)\right]=\sigma_{\rm Q}+\pi K \delta(\omega).\\
\end{split}
\end{equation}

We compute the Green's function by the standard holographic techniques that involve the solution of the EOM's corresponding to the perturbations to the metric $g_{tx}$ and to $a_x$. By using the bulk EOM's it is possible to express the equation for the fluctuation $a_x$ as a function of the bulk fields only. Using the solution of $a_x$ together with the bulk fields close to the boundary it is possible to write down the renormalized boundary action, which according to the usual holographic dictionary is related to the current.  The current-current Green's function is finally obtained by functional differentiation of the action. 
   
We now revisit \cite{jain2010,chakrabarti2011,Davison2015a} the calculation of the Drude weight in the case of an Einstein-Maxwell-dilaton model with a Lagrangian, 
\be\label{eq:action}
{\mathcal L} = R - \frac{Z(\phi)}{4} F_{\mu\nu}F^{\mu\nu} -{1\over2}\partial _\mu\phi\partial^\mu\phi+V(\phi)\ ,
\ee
which includes a non-minimal electromagnetic coupling $Z$ that may depend on the dilaton. The potential $V$ satisfies $V(\phi=0)=-2\Lambda$, where $\Lambda$ is the cosmological constant. For a detailed treatment of this model we refer to \cite{charmousis2010}. The conditions for the universal 
results of \cite{jain2010,chakrabarti2011,Davison2015a} are that the gauge field has no mass-terms and the boundary is still $AdS$. 

We assume that solutions of the EOM's only depend on the radial coordinate, $u=r_0/r$ ($r_0$ is the outer horizon), and $A_t(u) = A(u)$. The equation of motion of the fluctuation $\delta A_x=a_xe^{-i\omega t}$ is given by,
\be\label{eq:ax}
\frac{1}{\sqrt{-\gtt\guu}} \left(\sqrt{\frac{-\gtt}{\guu}} \gxx^{d-3\over2}Z a_x' \right)'  + \frac{Z\omega^2\gxx^{d-3\over 2}}{-\gtt }a_x= \left(\frac{Z^2 \gxx^{d-3\over2} A'^2}{-\gtt\guu} \right) a_x \,.
\ee
We stress that this equation is only valid  assuming that there is no vector potential mass terms in eq. (\ref{eq:action}). The $\omega^2$ term is needed to have consistent boundary conditions, though it does not enter in the calculation of the DC-conductivity. The Maxwell coupling is assumed to satisfy $Z \to 1 (Z_+)$ at $u=0$ ($u=1$), where $Z_+$ is determined by the value of the dilaton at the horizon. The component $\gtt$ ($\guu$) is assumed to have a single zero (pole) at the horizon and to be consistent with an asymptotically $AdS$ geometry. In other words, we assume (summing over $n$) $\gtt=g_0(1-u)+g_n u^n$, $\guu=\tilde g_0(1-u)^{-1}+\tilde g_n u^n$, $n\geq-2$. The constants $g_0$ and $\tilde g_0$ are temperature dependent. We assume an $AdS$ boundary $\gxx\propto u^{-2}$ where the constant of proportionality may be written in terms of the entropy density; $A(u)$ must vanish at the horizon and close to the boundary $A(u) \simeq \mu - {\rho/ r_0^{d-2}} u^{d-2}+\dots$ with $\rho$ the charge density and $\mu$ the chemical potential of the dual field theory.

The boundary condition at the horizon is
\begin{equation}\label{ingoing}
a_x=e^{-{i\omega \over 4\pi T}\log(1-u)}\left[a_1+{\cal O}(1-u)\right]\ ,
\end{equation}
where the prefactor of the logarithm follows from the constants $g_0$ and $\tilde g_0$ in the ansatz of the metric. The sign in the exponential, together with the time dependence $\left(e^{-i\omega t}\right)$ determines the ingoing character at the horizon. For small frequency, the general solution consistent with this boundary condition is
\be \label{axx}
a_x = C_1 a_x^{(0)}(u) +C_2 i\omega a_x^{(1)}(u)\ ,
\ee
where, at $u=1$, $a_x^{(0)}$ is a regular everywhere and $a_x^{(1)}$ has a singularity at the horizon. Moreover, $C_2=C_1 Z^+ a_x^{0}(u=1)^2 \left(\frac{s}{4 \pi}\right)^{d-3\over d-1} $ and $C_1$ is undetermined. It is fixed by imposing the second boundary condition at the asymptotic $AdS$ boundary, 
\be	
a_x(u\to0)=a_x^{(0)}(u\to0)=a_0=1\ .
\ee
In order to use eq. (\ref{sigma:def}) we need the current-current retarded Green's function, which, as we mentioned earlier, is obtained from the boundary action of the Lagrangian eq. (\ref{eq:action}).  It is easy to see that the only term which contributes to the required Green's function is obtained by double differentiation of
\begin{equation}
\lim_{u\to0}\sqrt{-g}g^{xx}g^{uu}Z a_x'a_x\ ,
\end{equation}
with respect to the boundary value of $a_x$. We have omitted the integral over space dimensions in the boundary. Moreover, as discussed before, eq. (\ref{Dirac:form}), the Drude weight is given by the ${\cal O}(\omega^0)$ contribution of the Green's function. Therefore, in the previous equation we only need to use the solution, $a_x^{(0)}$, namely:
\begin{equation}\label{Drude:def}
K=\lim_{u\to0}\sqrt{-g}g^{xx}g^{uu}Z {a_x^{(0)}}'\ .
\end{equation}
As we mentioned before, $a_x^{(0)}$ is regular in the whole domain. Therefore, we take $a_x^{(0)}=\sum_n a_n u^n$, $n\geq0$ with $a_0=1$ (normalization of the electric field). We now expand eq. (\ref{eq:ax}) with $\omega \to 0$ close to the boundary using the asymptotic form of $A_t$ together with the ansatz for $a_x^{(0)}$ and $g_{\mu\nu}$. This imposes constraints on the coefficients of $a_n$, which leads to
\begin{equation}\label{ax0:sol}
a_x^{(0)}=1-{\rho^2\over {d\over d-1}\epsilon}u^{d-2}+\dots\ ,
\end{equation}
where we used that the energy density enters through the expansion of $\gtt=g_0(1-u)+g_n u^n$, $g_{d-2}=-(d-1)\epsilon$. From eqs. (\ref{ax0:sol}) and (\ref{Drude:def}), it follows that the Drude weight agrees with the result derived in \cite{jain2010,chakrabarti2011,Davison2015a},
\begin{equation}\label{Kuniv}
K_{\rm U}={\rho^2\over \epsilon+P}\ ,
\end{equation}
where $\epsilon+P={d\over d-1}\epsilon$.

In the next sections we explore in more detail the limitations and extensions of the universal result $K_{\rm U}$ in several gravity backgrounds, including one with a vector potential mass term. 

For the moment we just comment the effect of a mass term $W A_\mu A^\mu$ in the Lagrangian (\ref{eq:action}). As we comment in sec. \ref{sec:deviation}, in order to avoid divergences, $W$ and its first derivative close to the boundary must tend to zero. Therefore, $W\propto u^n+\dots$, for $u\to0$, where the power and constant of proportionality depend on the boundary conditions of the dilaton. This mass term modifies eq. (\ref{eq:ax}) as well as the constraints on the coefficients of the ansatz of $a_x^{(0)}$, $a_n$. The new constraints yield an extra term $\OO(u^{d-2})$ in eq. (\ref{ax0:sol}). Therefore, in the presence of a massive vector potential, the Drude weight is in general different from the universal expression (\ref{Kuniv}).

Finally, we turn briefly to the temperature dependence of the universal Drude weight eq. (\ref{Kuniv}). 
In the canonical ensemble at least, it is expected $K_{\rm U}$ not to scale with temperature in the low temperature limit, since $\rho$ is fixed and the denominator is temperature independent, \cite{Davison2015}, which is consistent with our numerical results (not shown).

In very specific cases, such as a dimensionless charge density or chemical potential \footnote{Both are forbidden to be dimensionless simultaneously by the Gubser criterion, \cite{Kiritsis2015}.}, the temperature scaling in the low temperature limit may be obtained from simple dimensional analysis. 
The dimensionality of the relevant thermodynamic quantities are, $[\rho] = \tilde d - \theta + \Phi$, $[\mu] = z - \Phi$, $[s]=\tilde d- \theta$, $[T] =z$, where $\tilde d=d-1$ is the spatial dimension of the boundary, $z$ is the dynamical critical exponent, $\theta \neq 0$ is a signature of hyperscaling violations, and $\Phi$ is another critical exponent that controls the scaling of the gauge field around the horizon. 
For dimensionless chemical potential, $\Phi=z$ and $K\sim T^{\tilde d -\theta+z\over z}$ while for dimensionless charge density, $\Phi=\theta-\tilde d$ and $K\sim T^{-{\tilde d -\theta+z\over z}}$. We stress this is the prediction of dimensional analysis, which will be correct provided the dimensions of the chemical potential and charge density are not given by any other scale but the temperature.
In other cases an explicit numerical calculation is required.

\section{Universality of the Drude weight in theories with multiple massless gauge fields}\label{sec:Universiality2}
In this section we investigate the Drude weight in theories with several massless gauge fields.
The finite part of the DC conductivity in the models we discuss, but not the Drude weight, was investigated in detail in \cite{jain2009,jain2010,jain2010a}. We aim to clarify to what extent the universal results of the previous section can be extended to actions with multiple gauge fields. For that purpose we start with an action in $d+1$ bulk dimensions that is the natural generalization of eq. (\ref{eq:action}),
\begin{equation}
 {{\cal L} }=
 R  - {1\over 4} \sum_{i} Z_{i} F_{\mu\nu}^i 
 F^{\mu \nu\, i} + \ldots\ ,
\label{lagrangian}
\end{equation}
where $\ldots$ stand for scalar-fields or Chern-Simons terms.  At this stage it is not necessary to specify them since the calculation of the Drude weight involves solving the equation of the fluctuations of $A_x$, for which
it is not necessary to consider the fluctuations of the scalar fields. We only assume that these scalars do not condensate in the boundary. The extra index ($i$) in the Maxwell tensor $F_{\mu\nu}^i$, with strength coupling $Z_{i}$ that may depend of the scalar field, labels the $i$-th $ U(1) $ gauge field $A_{\mu i}$ of the theory. 
The equations of motion for the perturbations ${\delta A}_{xi}= A_{xi}(u)e^{-i\omega t + i q z}$ that control the conductivity, are simply,  see \cite{jain2009,jain2010} for details, 
\begin{equation}\label{axeq}
\frac{d}{du}\left(N_i\frac{d}{du}A_{xi}(u)\right)+\sum\limits_{j=1}^m  M_{ij}A_{xj}(u)+{\cal O}(\omega^2)=0\ ,
\end{equation}   
where the perturbation in the metric $\delta g_{xt}$, decouples from the equations of $A_{xi}$. We have omitted the term $-\omega^2\ N_i~ g_{uu} g^{tt}A_{xi}(u)$ since it is not needed to study the Drude weight.  The factors $N_i$ and $M_{ij}$ are (with no summation convention in $i,j$)
\begin{equation}
N_i=\sqrt{-g}Z_{ii}g^{xx}g^{uu},\quad  M_{ij}=F_{ut}^i\sqrt{-g}Z_{ii}g^{xx}g^{uu}g^{tt} Z_{jj}F_{ut}^j\ .
\end{equation}

As was shown in  \cite{jain2009,jain2010}, the regularized action at $u=u_c$ close to the boundary, necessary for the calculation of the conductivity is simply,
\begin{equation}\label{actmulti}
S_{u_c} =\frac {1}{16\pi G_{d+1}} \int \frac{d^{d}q}{(2\pi)^d}\left(\sum\limits_{i=1}^m N_i(u_c)\left.\frac{d}{du}A_{xi}(u,\omega,q)\right|_{u_c}A_{xi}(u_c,-\omega,-q)\right).
\end{equation}

The general expression for the Drude weight $K_{ij}$ is then obtained by functional differentiation of the boundary action, 
\begin{equation}
\begin{split}
&\hspace{1.5cm}K_{ij}=-\lim_{u_c\to0}\lim_{\omega, q\to 0}\text{Re}\left( G^R_{J^iJ^j}(\omega,q)- G^R_{J^iJ^j}(\omega=0,q)\right),\\
&G^R_{J^iJ^j}(\omega,q)=\frac {2}{16\pi G_{d+1}}\Bigg[\sum\limits_{k=1}^m N_k(u_c)\frac{\delta^2}{\delta A_{xi}^{(0)}\delta A_{xj}^{(0)}}{A_{xk}}'(u_c,\omega,q)A_{xk}(u_c,-\omega,-q)\Bigg],\\
\end{split}
\end{equation}
where $A_{xi}^{(0)}$ is the value of $A_{xi}$ at the boundary ($u_c=0$). Even before any explicit calculation of the conductivity is done, the above expressions suggest already several interesting features of the Drude weight in the multicharge case. It is clear that it is a tensor, namely, a small electric field related to the $i$ gauge field induces, in general, a current not only of the $i$ but also of the $j$ charge. This is a consequence of the non-linearity of the bulk equations. 

Moreover, as in the case of a single gauge field, the Drude weight is still exclusively controlled by the regular (no singularity around the horizon) solution. 
Since for a single charge the regular solution is $A_x = a_0 + c u^{d-2}$, for simple cases where $A_t$ is known explicitly, and eq. (\ref{axeq}) is linear we expect that the solution of eq. (\ref{axeq}) is given by
\begin{equation}
A_{xi} = a_0^i + u^{d-2} f(a_0^j,\rho_j,T\ldots)\ ,
\end{equation}
where $f$ depends, likely linearly, on $a_0^i$ and the rest of values of gauge fields at the boundary and other parameters such as temperature or the charge densities. On physical grounds $K_{ij}$ must be symmetric and in the limit of one charge must reproduce the universal result of previous section. Moreover, the linearity of the equations suggests that off-diagonal terms should not depend on powers of the charge density larger than two. The simplest expression for the Drude weight that meets these requirements is, 
\begin{equation}\label{multid}
K_{ij} \propto\frac{\rho_i \rho_j}{\epsilon+P}
\end{equation}

We now study in detail an example where the Drude weight is of the form given in eq. (\ref{multid}). This is a strong indication that this is the universal form of the Drude weight, eq. (\ref{Kuniv}), for the case of multicharges associated with massless gauge fields assuming AdS geometry in the boundary and no scalar-condensation.

Instead of embarking in numerical simulations with several gauge fields we will focus on a class of systems, R-charged backgrounds, where explicit analytical are available even for multicharges. Moreover, the field theory duals of these models are well known as these backgrounds come directly from compactifications of string theory. More specifically, we study the five dimensional R-charged black hole, also referred to as STU black 
holes \cite{Behrndt1996,Behrndt1999},
 whose field theory dual is a ${\cal N} = 2$ super Yang-Mills theory coming from the compactification of ten 
dimensional IIB supergravity on $S^5$, see \cite{Cvetic1999} for other cases involving the reduction of $D=11$ supergravity on $S^7$ and $S^4$. 
The action is given by
\begin{equation}
S = \frac{1}{16 \pi G} \int  d^5x \sqrt{-g} \Big( R + \frac{2}{l^2}{\cal V} + 
\frac{1}{2} 
G_{ij} F_{\mu\nu}^i
F_{\mu\nu}^j - G_{ij} \partial_\mu X^i \partial^\mu X^j + \frac{1}{24 {\sqrt{-g}}}
\epsilon^{\mu\nu\rho\sigma\lambda} \epsilon_{ijk} F_{\mu\nu}^i F^{\rho\sigma j} 
A_\lambda^k\Big),
\label{gaugeac}
\end{equation}
where $l$ represents the scale associated with the cosmological constant.
In addition to the metric, we have three scalar fields $X^i$, $i= 1,2,3$ while the scalar potential ${\cal V}$ and the metric $G_{ij}$ are given in terms of the scalar fields,
\begin{equation}
{\cal V} = 2 \sum_{1}^{3} \frac{1}{X^i},~~G_{ij} = \frac{1}{2} {\rm {diag}}\Big[ (X^1)^{-2}, (X^1)^{-2}, (X^1)^{-2}\Big]\ .
\end{equation} 
$F_{\mu\nu}^i$, $i=1,2,3$, are the field-strengths of  $A^i$, the Abelian gauge fields.

As shown in  \cite{Behrndt1996}, this effective action (\ref{gaugeac}) 
admits asymptotically AdS black hole solutions with three $U(1)$ charges. 
These solutions can be written down using the outer horizon $r_+$, the variable $u=r_+^2/r^2$ and $T_0={r_+\over\pi L^2}$ as 
\begin{equation}
ds^2 = -{\cal H}^{-2/3}(\pi L T_0)^2 {f(u)\over u} dt^2 + {\cal 
H}^{1/3}{L^2\over 4f(u)u^2} du^2 + {\cal H}^{1/3}{(\pi L T_0)^2 \over u}
(dx^2 + dy^2 + dz^2),
\end{equation}
where
\begin{equation}\label{R:charges:metric}
H_i = (1+k_iu),\quad i=1,2,3,\quad
{\cal H} = H_1 H_2 H_3\quad
f(u) = {\cal H} - \Pi u^2,\quad {\Pi}=\prod_{i=1}^3(1+k_i)\ .
\end{equation} 
The perturbed equations are given by,
\begin{equation}\label{Ax:eoms:R}
\begin{split}
A_{xj}'' + \left( \frac{f'}{f} - \frac{\cal H^\prime}{\cal H} + 2 
\frac{H_j'}{H_j} \right) A_{xj}' + 
\frac{{\tilde\omega}^2 {\cal H}} {u f^2}A_{xj}- u{\Pi\sqrt{k_j}\over f H_j^2}\sum_{i=1}^3\sqrt{k_i}A_{xi}=0\ ,\quad j=1,2,3\ ,
\end{split}
\end{equation} 
with $\tilde \omega={\omega\over 2\pi T_0}$. Following \cite{jain2010,jain2010a} we propose the following ansatz which satisfies the ingoing boundary condition,
\begin{equation}
A_{xi} = \dfrac{ f^{-i\tilde\omega (T_0/2T)}}{1 + k_i u} a_i(u)\ ,
\quad\quad i = 1,2,3\ .
\end{equation}
Since we aim to compute the Drude weight it is only necessary to expand $a_i$ up to leading order in $\tilde \omega$, 
\begin{equation}
a_i(u) =  \left[ a_i^{0}(u) + i\tilde\omega a_i^1(u) + {\cal O}({\tilde\omega}^2) \right],
\end{equation}
where, as before, the Drude weight tensor is extracted from $a_i^0$ only while for the real part of the conductivity $a_i^1$ is also needed. The equations of $a_i^0$ are simply,
\begin{equation}\label{a0:eom:R}
{a_j^0}'' + {a_j^0}'\left({f'\over f} - {{\cal H}'\over {\cal H}}\right)  + a_j^0{H_j'\over H_j}\left({{\cal H}'\over {\cal H}} - {f'\over f}\right) -   u{\Pi\sqrt{k_j}\over f H_j}\sum_{i=1}^3{\sqrt{k_i}a_i^0\over H_i}=0\ ,\quad j=1,2,3\ .
\end{equation}
A regular solution is easily found by substituting $a_i^0(u)=b_{i}+\tilde b_{i} u$ and solving the constraints resulting from the equations of motion eq. (\ref{a0:eom:R}). In this way $\tilde b_{i}$ is expressed as a function of the boundary values $b_{i}$ by
\begin{equation}
\begin{aligned}
&\tilde b_{j}={b_{j}\over2}k_j-\sum_{i\neq j}{b_{i}\over2}\sqrt{k_jk_i}
\end{aligned}
\quad \implies\quad
\begin{aligned}
&a_j^0(u)=b_{j}\left(1+{k_j\over2}u\right)-\sum_{i\neq j}b_{i}{\sqrt{k_jk_i}\over2}u\ .
\end{aligned}
\end{equation}

We now have extracted all the information of the gauge fields required to compute the Drude weigh. The part of the boundary action that contributes to the Drude weight, eq. (\ref{actmulti}), is
\begin{equation}
\begin{split}
\text{S}_{\rm{boundary}} &=\lim_{u\to 0}\frac{-r_+^2}{16\pi G L}\int ~ dtd\vec{x}\left[ {a^0_1}' a^0_1 + {a^0_2}' a^0_2 +\dots\right]=\\
&=\frac{r_+^2}{16\pi G L}\int ~ dtd\vec{x}\sum_{j=1}^3-\left(b_{j}{k_j\over2}-\sum_{i\neq j}b_{i},{\sqrt{k_jk_i}\over2}\right)b_{j}+\dots\ ,\\
\end{split}
\end{equation}
which leads to
\begin{equation}
K={1\over 16\pi G L}(-1)^{i+j}\sqrt{k_ik_j}r_+^2\ .
\end{equation}
In order to check the universality of this result it is illuminating to express the charges in terms of thermodynamic quantities, \cite{Son2006}.
The relevant thermodynamic quantities are given by,
\begin{equation}
\epsilon = {3\pi^2T_0^4N^2\Pi\over8},
\quad
P= {\epsilon\over3},\quad
\rho_i={\pi^2T_0^3N^2 2\sqrt{k_i}\sqrt{\Pi}\over8},
\end{equation} 
where $2G N^2=\pi L^3$ and $\Pi$ is given in eq. (\ref{R:charges:metric}). With these definitions the Drude weight can be expressed in terms of thermodynamic quantities, 
\begin{equation}\label{Drudemulti}
K_{ij} =  (-1)^{i+j}\frac{\rho_i\rho_j}{\epsilon+P}.
\end{equation}
Note that the off-diagonal terms are negative. The same occurs for the finite part of the DC conductivity \cite{jain2010a}. We do not yet have a clear physical interpretation of this feature. Obviously these prefactors cannot be universal as can be modified by a linear recombination of the currents. Only the eigenvalues of $K_{ij}$ are basis invariant. Because of this and the linearity of the equations leading to $K_{ij}$, we expect that, up to basis dependent prefactors, the above form of the Drude weight is likely universal for theories with several massless gauge field.

\section{Mazur-Suzuki bounds and holographic correlation functions}
In this section we introduce the so called
Mazur-Suzuki (MS) bounds \cite{mazur1969,suzuki1971,zotos1997}, inequalities among correlation functions that describe transport in interacting many-body problems. We then discuss how these correlation functions are expressed in terms of holographic retarded Green's functions and relate them to the Drude weight studied in previous sections. We shall see, by working out some examples in Einstein-Maxwell and R-charged backgrounds, that the correlation functions are not given entirely by the zero-momentum retarded Green's functions obtained with the standard recipe in holography.

The main result of the section is that, in the models we study, the MS bound is saturated only if the Drude weight is given by the universal result (\ref{Kuniv}).

As we mention previously a finite Drude weight is a signature of ballistic non-dissipative transport. Indeed Kohn \cite{kohn1964} proposed to characterize non-disordered metals and insulators attending to whether the Drude weight was finite or not respectively. This non-dissipative transport must be caused by the non-decay of certain correlation functions, in this case the electrical current-current correlation. It is well known that the existence of conservation laws can protect the decay of certain correlation functions. 
This was precisely the starting point of Mazur analysis that we briefly review next. Let us consider all the conserved quantities $Q_i'$ of the system, namely, $[H,Q'_i]=0$, $[Q'_i,Q'_j]=0$. By some rearrangements, it is possible to chose them orthogonal each other $\langle Q_i Q_j \rangle = Q_i^2\delta_{ij}$. 
An operator, the electric current in our case, can be expressed in terms of these conserved quantities:
\begin{equation}\label{Mazur}
K = {\beta\over V}\lim_{t \to \infty} \langle J(t)J(0) \rangle = \lim_{N \to \infty}{\beta\over V} \sum_i^N \frac{\langle  JQ_i\rangle^2}{\langle Q_iQ_i\rangle}\ ,
\end{equation}
where $\beta$ is the inverse of the temperature and $V$ the volume. The correlation functions on the right-hand side are for large times. Since each term in the right hand side is positive,
\begin{equation}\label{suz}
K \geq K_{\rm MS}\equiv {\beta\over V}\sum_i^k \frac{\langle  JQ_i\rangle^2}{\langle Q_iQ_i \rangle},~ k<\infty \ .
\end{equation}
$K_{\rm MS}$ is the Mazur bound for the Drude weight, $K$, first obtained in Refs. \cite{zotos1997,zotos1999}. Its generalization to other operators is straightforward. 

We stress that the inequality is usually more useful than the equality since, by picking up a single conserved quantity, it allows to find out, at least in some cases, whether the Drude weight is finite or not. For instance in strongly interacting one-dimensional systems an explicit calculation of the Drude weight is typically very demanding while the calculation of the right hand side, for instance for the energy current which sometimes is a conserved quantity, is much easier as it involves only static correlation functions. In the following sections  we compute the MS bound in the following gravity duals: the Einstein-Maxwell theory with a without a scalar that induces $U(1)$ symmetry breaking and a R-charged background where explicit solutions for the background metric are available. For that purpose we will have first to express the bound in terms of susceptibilities, namely, retarded Green's functions which is the natural language in holography. This is indeed the way that Suzuki \cite{suzuki1971} proceeded to extend the classical Mazur bounds to quantum mechanical systems.

\subsection{Mazur-Suzuki bounds in Einstein-Maxwell theory}

We start our analysis with the Einstein-Maxwell theory,

\begin{equation}\label{action}
\begin{split}
S=&-{1\over 2\kappa^2}\int_{\cal M} d^{d+1}x\sqrt{-g}\left(R+{d(d-1)\over L^2}+{1\over 4e^2}F_{\mu\nu}F^{\mu\nu}\right)+\\
&-{1\over 2\kappa^2}\int_{ \partial{\cal M}} d^dx\sqrt{-\tilde \gamma}\left(-2K+2{d-1\over L}\right)\ ,\\
\end{split}
\end{equation}
where $\tilde\gamma$ is the boundary metric induced by $g$ and $K$ is the trace of the extrinsic curvature. The last integral includes the counterterms needed to have a well defined energy tensor in the boundary. These counterterms include powers of the induced Ricci scalar on the boundary, but since $\cal M$ is asymptotically flat they do not contribute.
The solution of the EOM's is the AdS planar Reissner-Nordstr\"{o}m (RN) background in $d+1$ dimensions,
\begin{equation}\label{metric}
\begin{split}
&ds^{2}={1\over L^2 z^2}\left(-f(z) dt^2+{L^4\over f(z)}dz^2+dx_i^2\right),\\
& f(z)=1-(1+Q^2) {\left(z\over z_0\right)}^d+Q^2{\left(z\over z_0\right)}^{2d-2}\ ,\\
\end{split}
\end{equation}
where $i=1,\dots,d-1$, $z=1/r$. The only non-zero component of the gauge field is $A_t=\phi=\mu\left[1-(z/z_0)^{d-2}\right]$, $Q^2={\mu^2 z_0^{2}\gamma^{-2}}$, $\gamma^{-2}={d-2\over d-1}{L^4\over 2}$, $z_0$ is the inverse of the outer horizon and we set
\begin{equation}\label{kappa}
{2\kappa^2\over e^2}=1\ .
\end{equation}
In order to calculate the electrical conductivity in the linear response approximation we add a time-dependent weak perturbation in the gauge field and the metric, $A_x(z)e^{-i\omega t}$, $g_{tx}(z)e^{-i\omega t}$. The equations of motion of $A_x$ and $g_{xt}$ are:
\begin{equation}
\partial_z(fz^{3-d}A_x)+A_x\left({\omega^2z^{3-d}\over f}-\phi'^2z^{5-d}\right),\ \ g_{xt}'+{2\over z}g_{xt}+A_x\phi'=0\ .
\end{equation}
Close to the boundary they satisfy,
\begin{equation}
\begin{split}
&A_x\sim \axa+\axb {z\over z_0}^{d-2},\quad \gxt\sim {\gxta\over z^2}+\gxtb z^{d-2},\quad\gxtb=\mu{d-2\over d}{\axa\over z_0^{d-2}}\ ,\\
\end{split}
\end{equation}
where the prime denotes differentiation with respect to $z$ and $\gxta$ and $\axa$ source the operators dual to $A_x$ and $g_{xt}$.
\subsubsection{Calculation of the MS bounds}
Assuming that the conserved quantity is momentum, the MS bound depends on boundary momentum-momentum and current-momentum correlators. The evaluation of the on-shell action eq. (\ref{action}) on the boundary results in the following terms relevant for the calculation of the corresponding Green's functions,
\begin{equation}
\begin{split}
S=&{V_{d-1}\over 2\kappa^2 2\pi}\left[{(d-1)(1+Q^2)\over2 z_0^d}\gxta(-\omega)\gxta(\omega)+\right.\\
&\left.+{\mu(d-2)\over 2z_0^{d-2}}(\axa(\omega) \gxta(-\omega)+\axa(-\omega) \gxta(\omega))\right]+\dots\\
\end{split}
\end{equation}
where $V_{d-1}$ is the boundary spatial volume.
Therefore the retarded Green's functions at zero spatial momentum are,
\begin{equation}\label{green_fns}
\begin{split}
G_{J_x\Pi_x}(\omega)&={e(d-2)\mu\over 2\kappa^2 z_0^{d-2}}{V_{d-1}\over 2\pi}\ ,\\
G_{\Pi_x\Pi_x}(\omega)&={(d-1)(1+Q^2)\over 2\kappa^2 z_0^{d}}{V_{d-1}\over 2\pi}\ .\\
\end{split}
\end{equation}
This agrees with the results for $d=3,4$, available in \cite{Hartnoll2008} for holographic superconductors in the normal state, and in \cite{Matsuo2009} for a Reissner-Nordstr\"{o}m background after setting all the perturbations in the metric to zero, except $h_{zt}$, which corresponds in our notation to $\gxt$. We note however that the result in eq. (\ref{green_fns}) is not enough, in general, to obtain the  correlation functions that enter in the MS bounds. We now discuss the exact relation of the Green's functions and the correlation functions needed in the MS bound. 

For simplicity, let us consider a single conserved quantity $Q_1=Q$ and the conserved current $J$ in eqs. (\ref{Mazur}) and (\ref{suz}). As we mentioned in the beginning of the section the correlation functions in these equations are for large times, 
\begin{equation}\label{larget_corr}
\langle J Q\rangle\equiv \lim_{t\to\infty}\langle J(t) Q(0)\rangle~,\langle Q Q\rangle\equiv \lim_{t\to\infty}\langle Q(t) Q(0)\rangle~.
\end{equation}
In order to relate these correlation functions to Green's functions we introduce some standard notation in linear response theory, \cite{Forster,Mazenko}.  Consider the variation of an observable, $\delta \langle A_i({\bf r},t)\rangle$ due to external perturbations  $\delta \langle a_i^{ext}({\bf r},t)\rangle$. The {\it Kubo correlation function}, defined as,
\begin{equation}
\begin{split}
C_{ij}({\bf r},{\bf r}',t-t')=&{1\over\beta}\int_0^\beta d\lambda\langle e^{\lambda H}\delta A_i({\bf r},t)e^{-\lambda H}\delta A_j({\bf r}',t')\rangle~,\\
\end{split}
\end{equation}
where $H$ is the unperturbed Hamiltonian. The Laplace transform of the Kubo correlation function satisfies \cite{Forster,Mazenko},
\begin{equation}\label{Cijthm}
C_{ij}({\bf r},{\bf r}';z)={-1\over  \beta z}\left[\chi_{ij}({\bf r},{\bf r}',z)-\chi_{ij}({\bf r},{\bf r}',i0)\right]~,
\end{equation}
where $z$ is the transformed variable of $t$ and $\chi_{ij}$ is known  in the literature as the admittance, matrix response function as well as the Green's function. 

It is therefore natural to express the large time correlation functions  in eq. (\ref{larget_corr}) in terms of the low frequency limit of the retarded Green's functions as,
\begin{equation}\label{rel_corr_G}
\begin{split}
&\langle JQ\rangle={1\over \beta}\lim_{\omega\to0,q\to0}\left[G_{JQ}(\omega,q)-G_{JQ}(0,q)\right]\ ,\\
&\langle QQ\rangle={1\over \beta}\lim_{\omega\to0,q\to0}\left[G_{QQ}(\omega,q)-G_{QQ}(0,q)\right].
\end{split}
\end{equation}
With these definitions we have now all the information to compute the MS bound associated to the electrical conductivity in the Einstein-Maxwell theory. 

The MS bound relates the Drude weight $K$ with correlation functions between the current and conserved charges, see eq. (\ref{suz}). For the case of the electrical conductivity, $\sigma$, we use the following notation in  eq. (\ref{suz}), $V=V_{d-1}$ is the spatial volume on the boundary, $\beta=1/k_BT$, $J=J_x$ is the current associated to $\sigma$, $Q_j$ are the conserved charges which overlap with $J_x$ and $k$ stands for a certain number of conserved charges. If all possible conserved charges are considered the bound is saturated. 
In our system momentum is conserved so it is natural to set $k=1$ and $Q_1 = \Pi_x$, which in a relativistic field theory corresponds to the spatial components of the diagonal of the energy momentum tensor. With this identification the numerator of (\ref{suz}) is  given in terms of $\langle J_x\Pi_x\rangle$, which according to eq. (\ref{rel_corr_G}) is obtained from $G_{J_x\Pi_x}(\omega,q)$. However, due to the dependence of $G_{J_x\Pi_x}(\omega,q)$ on the frequency, \cite{Matsuo2009}, $G_{J_x\Pi}(0,q)=0$ and thus we may use the zero-momentum Green's function $G_{\Pi_xJ_x}$ given in eq. (\ref{green_fns}) 
\begin{equation}\label{JPRN}
\langle J_x\Pi_x\rangle={1\over\beta}\lim_{\omega\to0}G_{J_x\Pi_x}={1\over\beta}{e(d-2)\mu\over 2\kappa^2 z_0^{d-2}}{V_{d-1}\over 2\pi}~.
\end{equation}
However, this is not the case for the denominator, $\langle \Pi_x\Pi_x\rangle$, for which $G_{\Pi_x\Pi_x}(\omega=0,q)\neq0$, as seen in \cite{Matsuo2009} for $d=4$.  For arbitrary $d\geq3$ we cannot use the zero-momentum $G_{\Pi_x\Pi_x}$ given in eq. (\ref{green_fns}). Nonetheless,  $\langle \Pi_x\Pi_x\rangle$ is the momentum static susceptibility, which may be written in terms of hydrodynamical quantities, $\chi_0=\langle\epsilon + P\rangle$, $\epsilon$ and $P$ being the energy density and pressure \cite{Herzog2009}. An identical result is obtained by using Ward identities \cite{hoyos2015}. Therefore the momentum-momentum correlation function needed in the MS bound is in this case,
\begin{equation}\label{staticsuscept}
\langle\Pi_x\Pi_x\rangle=\chi_0=\langle\epsilon + P\rangle~.
\end{equation}
Finally, eqs. (\ref{suz}), (\ref{JPRN}) and (\ref{staticsuscept}) yield
\begin{equation}\label{K_hydro}
K(T)\geq K_{\rm MS}={\beta\over V_{d-1}}{\langle J_x\Pi_x\rangle^2\over \langle\Pi_x\Pi_x\rangle}={(d-2)^2\mu^2\over d(1+Q^2)}{z_0^{4-d}\over 2\kappa^2}~.
\end{equation}
The horizon $z_0$ depends on temperature through the standard relation for a RN black hole. We used that for the Einstein-Maxwell theory given by eqs. (\ref{action}) and (\ref{metric}), $\rho={(d-2)\mu z_0^{-d+2}\over e^2}$, $\epsilon=(d-1)P$, \cite{Kovtun2012} and  $\epsilon=z_0^{-d}(d-1)(1+Q^2)$, $Q=\mu z_0/\gamma$, defined above.

This result is to be compared the explicit calculation of the Drude weight $K(T)$ that yields the universal result
\cite{Hartnoll2007b} \cite{jain2009}\cite{Davison2015a}, 
\begin{equation}\label{hydroK}
K=K_{\rm U}=\frac{\rho^2}{\epsilon+P},
\end{equation}
where $\rho$, $\epsilon$ and $P$ are the charge and energy densities and the pressure, respectively. Substituting $\rho$, $P$ and $\epsilon$ in eq. (\ref{hydroK}), and setting $e=1$, it is clear that in this background the MS bound is saturated $K = K_{\rm MS}$.

 We note that in the condensed matter literature it is conjectured  \cite{castella1995,zotos1997,mierzejewski2014} that a finite Drude weight is a signature of integrability. In principle, this result is applicable to the field theory dual of the gravity action we investigate. In  classical gravity 
 In \cite{Giataganas2014} integrability of various gravity backgrounds has been recently studied. More specficially it was proposed a relation between integrability and saturation of the null energy conditions. That precludes integrability in most non-relativistic backgrounds. Integrability in four dimensional Einstein-Maxwell theory with a cosmological constant has been recently studied \cite{Klemm2015}. Clearly, further research is needed to understand to what extent a finite Drude weight might be a signature of integrability of a classical gravity theory and its dual field theory. For the moment we only comment that in the large $N$ limit there are drastic simplifications, even in QCD, in the dynamics of quantum field theories. Therefore, we cannot rule out that integrability plays a role in the occurrence of a finite Drude weight. 
 
\subsection{Mazur-Suzuki bounds in an R-charged black hole}
We now study another example where explicit analytical results for the background metric are known. 
We work with the 2- and 1-R-charged black holes in five dimensional $N=2$ $U(1)^3$ gauged supergravity, \cite{Dewolfe2011}, which are particular cases of the more general model studied in Ref. \cite{Maeda2008a}. They are obtained by setting  two of the three $U(1)$ charges to be equal, $Q_1=Q_2=Q\neq Q_3$. The 2-R-charged black hole corresponds to $Q_3=0$, while setting $Q_1=Q_2=Q=0$ is referred to as the 1-R-charged black hole, \cite{Dewolfe2011}.

In the 1-R-charged black hole, the conductivity,
\begin{equation}
\sigma={r_H^2\over L^3}{2A_x^{(1)}\over i\omega A_x^{(0)}}-i{\omega\over2}\ ,
\end{equation}
has been calculated perturbatively in \cite{Dewolfe2011}:
\begin{equation}\label{sigma:sugra2}
\sigma\sim i{Q^2\over 2\kappa^2\omega L}+{L(Q^2+2r_H^2)^2\over8r_H\kappa^2\sqrt{Q^2+r_H^2}}+\cal{O}(\omega)\ .
\end{equation}
The temperature and chemical potential, expressed in terms of the charge, $Q$, and horizon, $\rh$, of the black hole are:
\begin{equation}
T={Q^2+2\rh^2\over 2\pi L^2\sqrt{Q^2+\rh^2}}\,, \qquad \Omega={\rh Q\over L^2\sqrt{Q^2+\rh^2}}\ .
\end{equation}
We note that eq. (\ref{sigma:sugra2}) matches the universal result, eq. (\ref{Kuniv}). The Green's functions needed to calculate the MS bound have been given in \cite{Son2006}, which in the notation of \cite{Dewolfe2011}\footnote{There is a difference definition for the vector potential in \cite{Son2006}, which should be multiplied by $\sqrt{2}$ in the notation of \cite{Dewolfe2011}},
\begin{equation}
\begin{split}
&G_{\Pi_x,\Pi_x}(\omega,q)=-{2q^2\rh^2(Q^2+\rh^2)\over \kappa^2 L^3(L^2q^2-4\omega i\sqrt{Q^2+\rh^2})}V_3,\\
&G_{\Pi_x J_x}(\omega,q)={4 \rh \sqrt{Q^2 + \rh^2} \sqrt{
 Q^2 (Q^2 + \rh^2)} \omega \over L^3 \kappa^2 (i L^2 q^2 + 
   4\sqrt{Q^2 + \rh^2} \omega)}V_3,\\
   \end{split}
\end{equation}
where $q$ is the spatial momentum of the perturbations $A_x(r)e^{i\omega t+i q x}$ and $V_3$ is the spatial volume in the boundary.
With these considerations,
\begin{equation}\label{corr_Rcharged}
\begin{split}
&\langle \Pi_x J_x\rangle={1\over \beta}\lim_{\omega\to0,q\to0}\left[G_{\Pi_x J_x}(\omega,q)-G_{\Pi_x J_x}(0,q)\right]\ ,\\
&\langle \Pi_x \Pi_x\rangle={1\over \beta}\lim_{\omega\to0,q\to0}\left[G_{\Pi_x \Pi_x}(\omega,q)-G_{\Pi_x \Pi_x}(0,q)\right].
\end{split}
\end{equation}
Finally, the MS bound is obtained from eqs. (\ref{suz}) with a single conserved quantity $Q_1=\Pi_x$ and eq. (\ref{corr_Rcharged}),
\begin{equation}
K\geq{ Q^2\over 2\kappa^2 L}\ .
\end{equation}
Comparing the MS bound with the exact result, given by the $\omega^{-1}$ term in  eq. (\ref{sigma:sugra2}), we see the bound is again saturated and the Drude weight is still the universal one, eq. (\ref{Kuniv}). 

Similarly, the zero frequency conductivity for the 2-R-charged black hole has been calculated exactly, \cite{Dewolfe2011}, and is also given by the universal result.

Based on these examples, it seems that if a theory with gravity dual is well described by hydrodynamics, like those dual to asymptotically $AdS$ EMd  theories, the Drude weight is given by the universal result (\ref{hydroK}) and the MS bound is saturated. 

\subsection{Mazur-Suzuki bounds in $U(1)$ spontaneously broken symmetry backgrounds}
We found previously that the MS bound is saturated in asymptotically $AdS$ EMd backgrounds where the Drude weight $K = K_{\rm U} = { \rho^2 \over \epsilon +P}$. Here we compute the MS bound in Einstein-Maxwell-scalar theory, \cite{Hartnoll2008}, which displays a spontaneous $U(1)$ symmetry breaking due to scalar condensation. In this background, it has been shown, \cite{Hartnoll2008,Herzog2009}, that the Drude weight receives an extra contribution related to the superfluid density.

In order to compute the Green functions that enter the MS bound it is necessary to obtain the properly renormalized boundary action. We just state the main result and refer to \cite{Hartnoll2008} for details,
\begin{equation}\label{HoloSC:bdy:action}
\begin{split}
S=&{V_{d-1}\over 2\kappa^2 2\pi}\left[{(d-1)(1+Q^2)\over2 z_0^d}\gxta(-\omega)\gxta(\omega)+\right.\\
&\left.+{\mu(d-2)\over 2z_0^{d-2}}(\axa(\omega) \gxta(-\omega)+\axa(-\omega) \gxta(\omega)) \psi^{(0)}\psi^{(1)} \right]+\dots\\
\end{split}
\end{equation}
where $\psi^{(0)},\psi^{(1)}$ are the coefficients of the scalar expansion close to the boundary $\psi \sim \psi^{(0)}z+\psi^{(1)}z^2$. We note that the only difference with respect to the non-condensed case is the last term. Interestingly, for a $U(1)$ symmetry-breaking to be spontaneous,  either $\psi^{(0)}$, or $\psi^{(1)}$ must vanish (depending on the quantization). This implies that the last term in the boundary action in eq. (\ref{HoloSC:bdy:action}) does not contribute to the Green's functions $G_{\Pi_x,\Pi_x}$ and $G_{J_x,\Pi_x}$. As a consequence the MS bound coincides with the one with no $U(1)$ symmetry breaking and 
\begin{equation}
K_{\rm MS}={\beta\over V_{d-1}}{\langle J_x\Pi_x\rangle^2\over \langle \Pi_x\Pi_x\rangle}=\frac{\rho^2}{\epsilon + P}\ .
\end{equation}
However, the bound is not saturated because of the additional superfluid contribution so $K >K_{\rm MS} =  K_{\rm U}$.

It would be interesting to compute the MS bound in theories with double trace deformations where it possible to have spontaneous symmetry breaking with both $\psi^{(0)}$ and $\psi^{(1)}$ non-zero. In sec. \ref{sec:EMDW} we will investigate in more detail the extra contribution to the Drude weight on a more general background, an Einstein-Maxwell-dilaton background with a gauge mass term.

In the following sections we study the Drude weight and the MS bound in non-relativistic backgrounds: the Einstein-Proca and an asymptotically Lifshitz EMd model with two gauge fields. 
As before, the calculation requires a properly renormalized boundary action and a careful evaluation of the correlation functions.  We shall see that $G_{J_x,\Pi_x}$ vanishes in all cases so $K_{\rm MS }=0$ and the bound is trivial $K > 0$. Moreover, the Drude weight is not given by the universal result. This suggests that the bound is only saturated if the Drude weight is given by the universal expression.

\section{Deviations from universality I: non-relativistic boundary field theory} \label{sec:deviation}

One of the conditions for the universal result of the zero-frequency conductivity is that the metric approaches $AdS_d$ in the boundary. This a necessary condition for the dual field theory to be relativistically invariant. However, in recent years the potential for condensed matter applications, that are typically described by non relativistic theories,  have stimulated the interest in asymptotic non-AdS gravity backgrounds. There are different actions that lead to these types of background \cite{Kachru2008,Taylor2008,Balasubramanian2009}. Here, we compute the Drude weight for the case of an EMd action with two gauge fields, \cite{Tarrio2011}, and for an Einstein-Proca action, which involves a massive gauge field \cite{Taylor2008}. A way to break relativistic invariance in the boundary is by 
imposing that after a change of scale $\lambda$, the time and space coordinates scale differently, $t \to \lambda^z t$, $x^i \to \lambda x^i$,
where $z \geq 1$ is the dynamical critical exponent. The simplest metric with this symmetry is,
\begin{equation}\label{metric:Lifshitz:zeroT}
ds^{2}=-{r^{2z}\over L^{2z}}dt^{2}+\frac{L^2}{r^{2}}dr^{2}+{r^{2}\over L^2}\sum\limits^{d}_{i=1}dx^{2}_{i}.
\end{equation}
\subsection{Asymptotically Lifshitz EMd model}

We start with the case of an Einstein-Maxwell-dilaton action with two (massless) vector fields:
\begin{equation}\label{S:Lifshitz}
S=\frac{1}{16\pi G_{4}}\int
d^{4}x\sqrt{-g}\left(R-2\Lambda-{1\over2}\partial_\mu\phi\partial^\nu\phi-\frac{e^{\lambda_1 \phi}}{4}F^{2}-\frac{e^{\lambda_2 \phi}}{4}G^{2}\right),
\end{equation}
$F=dA$, $G=dB$ and $\Lambda=-{(z+2)(z+1)\over 2L^2}$. The solution is, \cite{Tarrio2011}
\begin{equation}
ds^{2}=-{r_h^{2z}\over L^{2z}}f(r)dt^{2}+\frac{L^{2}}{r^{2}f(r)}dr^{2}+{r^2\over L^2}(dx^{2}+dy^2)\ ,
\end{equation}
with $z\geq1$ and
\begin{equation}\label{fields:Lif}
\begin{split}
&\phi=\log{\phi_0r^{\sqrt{4(z-1)}}},\ \ f=1-\left(1+{\rho_2^2\over \phi_0^{\sqrt{z-1}}4z}\right)\left({r_h\over r}\right)^{z+1}+{\rho_2^2 L^{2z}\over \phi_0^{\sqrt{z-1}}4z}{1\over r^{z+1}}\ ,\\
& A_{t}={\sqrt{2(z-1)(z+2)}\over L^z(z+1)}\left(r^{2+z}-r_h^{2+z}\right),\ \ B_{t}={\rho_2\phi_0^{-\sqrt{z-1}}\over z}\left(r_h^{-z}-r^{-z}\right).
\end{split}
\end{equation}
The gauge field with divergent time component {\it supports} the asymptotically Lifshitz geometry and does not contribute to the thermodynamic properties of the boundary theory, \cite{Tarrio2011}. The charge density of the second gauge field, $B_\mu$, is
\begin{equation}\label{Lif:charge}
q_2={L^{z-1}\over 16\pi G_4}\rho_2\ ,
\end{equation}
while its boundary value may be read from eq. (\ref{fields:Lif}). The entropy density and temperature are as follow
\begin{equation}
s={r_h^{2}\over 4G_4},\ \ T= {r_h^z\over 4\pi L^{z+1}}\left[z+2-{\rho_2^2 L^{2z}\over \phi_0^{\sqrt{z-1}}4  r_h^{2z+2}}\right].
\end{equation}
The boundary theory of eq. (\ref{S:Lifshitz}) is renormalized by adding the following counterterms, \cite{Dehghani2015}
\begin{equation}\label{ct:Lif}
S_{ct}={1\over 16\pi G_4}\int d^3x\sqrt{-\gamma}\left(2K-{4\over L}+c_A\sqrt{-e^{\lambda_1\phi} \gamma^{ij} A_iA_j}\right),\ \ c_A=-\sqrt{2(z-1)(2+z)}\ ,
\end{equation}
where $\gamma_{ij}$ is the induced metric in the boundary, $\gamma$ its determinant  and $K=\gamma^{\mu\nu}\nabla_\mu n_\nu$, $n_\mu$ is normal to the boundary and points outward. See \cite{Dehghani2015} for a more general model with a hyperscaling violation exponent. It is easy to check the renormalized boundary action given by eqs. (\ref{S:Lifshitz}) and (\ref{ct:Lif}) give the correct result for the Gibbs thermodynamical potential. The following term
\begin{equation}\label{canonical}
S_{canonical}={1\over 16\pi G_4}\int d^3x \sqrt{-\gamma}e^{\lambda_2\phi}n_\mu G^{\mu\nu}B_\nu,
\end{equation}
should be added to obtain the Helmholtz free energy from the action. 

\begin{figure}[H]
\center
\includegraphics[scale=0.62,clip]{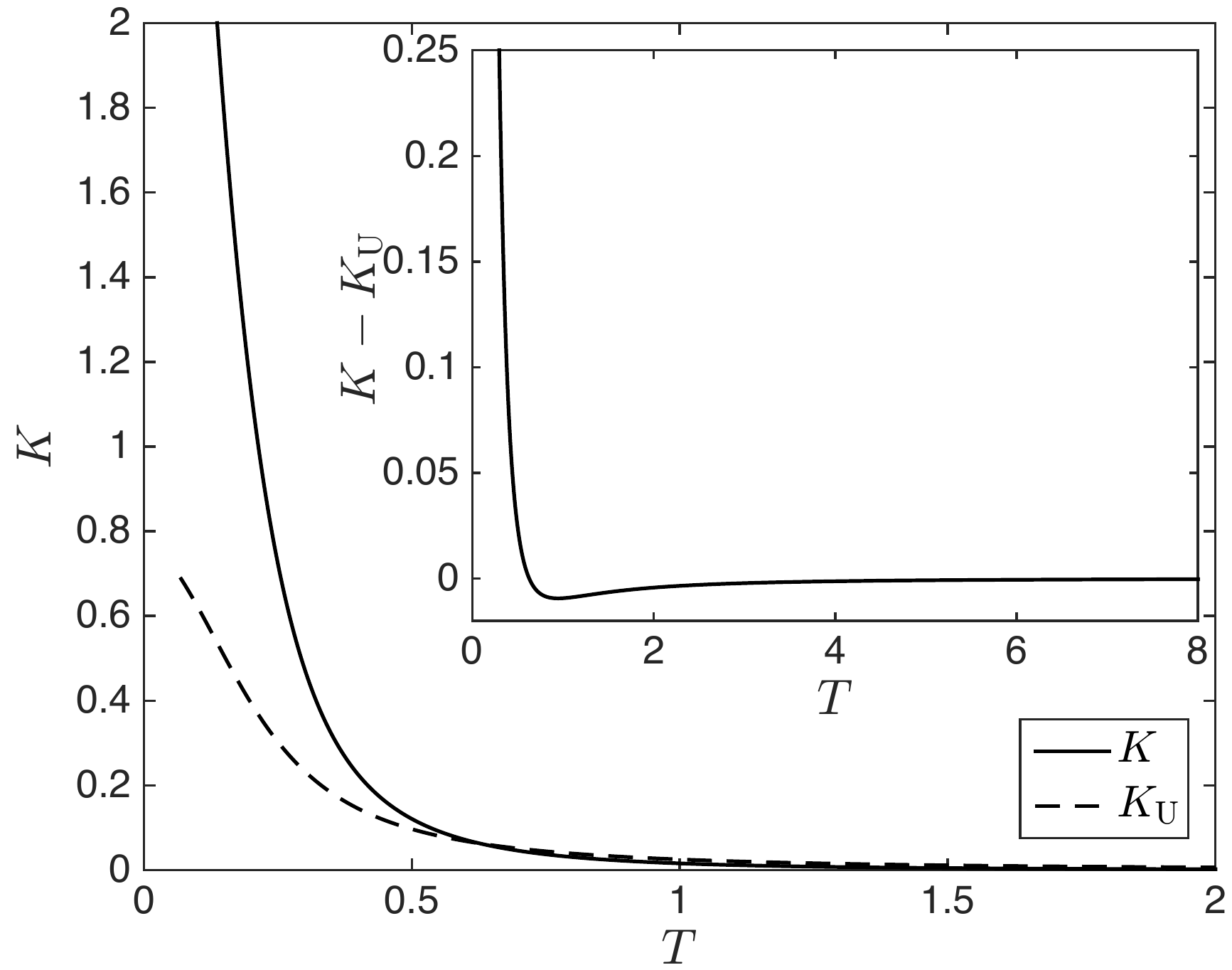}
\vspace{-0.4cm}
\caption{Drude weight in the asymptotically Lifshitz model, eq. (\ref{S:Lifshitz}) with $z=2$. We have fixed the charge density $q_2=1$, $\phi_0=1$ in units of $L=1$. The dashed line is the universal result, $K_{\rm U}={q_2^2\over\epsilon+P}$, eq. (\ref{Kuniv}) with the charge density $q_2$ given in eq. (\ref{Lif:charge}). The inset figure shows that the Drude weight is always different from the universal prediction $K_{\rm U}$. Interestingly the difference is non-monotonic. For low temperatures $K > K_{\rm U}$ while in the high temperature limit $K > K_{\rm U}$ since $K_{\rm U}\sim T^{-2}$ and $K\sim T^{-3}$. Moreover the MS bound vanishes $K_{\rm MS}=0$ so there is no saturation $K > K_{\rm MS} = 0$. }
\label{K_Lif}
\end{figure}

In order to study the zero frequency conductivity we add the perturbations
\begin{equation}
\delta g_{xt}=\tilde g_{xt}e^{-i\omega t},~~\delta B_x= \tilde B_x e^{-i\omega t},
\end{equation}
which satisfy,
\begin{equation}\label{Bx_eom_Lif}
\tBx''+\tBx'\left({f'\over f}+{z+1+r\lambda_2\phi'\over r}\right)-\tBx\left( {e^{\lambda_2\phi}  L^{2z}B_t'^2\over fr^{2z}}+\omega^2{L^{2z+2} \over f^2 r^{2z+2}}\right)=0,\quad \tgxt-{2\over r}\tgxt +e^{\lambda_2\phi}\tBx B_t'=0\ .
\end{equation}
We impose ingoing boundary conditions at the horizon,
\begin{equation}
\tilde B_x\simeq f(r)^{-i{\omega\over 4\pi T}}\left(b^{(0)}_x(r)+\omega b^{(1)}_x(r)+\dots\right),
\end{equation}
and solve for $\tilde g_{xt}$ and $\tilde B_{x}$ perturbatively in frequency. To obtain the Drude weight we only need to find $b^{(0)}_x(r)$. We have not been able to get  an analytical solution for $b^{(0)}_x$. However, it is easy to solve eq. (\ref{Bx_eom_Lif}) numerically. As is observed in Fig. \ref{K_Lif}, the Drude weight is finite but it is not given by the universal result $q_2^2\over \epsilon+P$. Moreover, by computing the boundary action explicitly, it follows the electric current dual to $\tilde B_x$ does not couple to the momentum. Therefore the MS bound is of not relevance and it always vanishes $K > K_{\rm MS} = 0$.

\subsection{Asymptotically Lifshitz Einstein-Proca model}

It has recently been shown,  \cite{Taylor2008}, that the metric given in eq. (\ref{metric:Lifshitz:zeroT}) is also the solution of an Einstein-Proca action, which includes a massive gauge field and gravity with a negative cosmological constant. The renormalization of this theory has been extensively studied, \cite{Ross2009,Ross2011,Mann2011,Baggio2012}. In \cite{Griffin2012,Baggio2012b}, an additional bulk scalar has also been included in the action. Finally, a comprehensive formalism to study the dual theory to
\begin{equation}
S={1\over 2\kappa^2}\int d^{d+1}x \sqrt{-g}\left(R-\alpha (\partial \phi)^2- Z(\phi)F^2-W(\phi')A^2-V(\phi)\right)
\end{equation}
has been introduced in \cite{Chemissany2015}, including the corresponding counterterms. An asymptotically Lifshitz background at finite temperature is obtained if there are two gauge fields $F=dA,~F_1=dB$ with only one being massive. More explicitly the action in this case is
\begin{equation}\label{S:Lifshitz:mass}
S=\frac{1}{16\pi G_{d+2}}\int
d^{d+2}x\sqrt{-g}(R-2\Lambda-\frac{1}{4}F^{2}-\frac{1}{2}m^{2}A^{2}-\frac{1}{4}F_{1}^{2})\ ,\end{equation} 
where the dynamical critical exponent is fixed to $z=2d$. It is also possible to solve analytically the perturbations needed to compute the electrical conductivity at zero frequency. However, the proper renormalization of the action eq. (\ref{S:Lifshitz:mass}) has not been settled. Therefore, the Drude weight or the MS bound of the dual theory cannot yet be computed rigorously. 

For that reason we study the simpler model introduced in Refs. \cite{Korovin2013,Korovin2013b} consisting on bulk gravity coupled to a single massive vector field.  
 
Since we are interested in finite temperature solutions we focus on the action studied in \cite{Korovin2013b},
\begin{equation}\label{S:Skenderis}
S=\frac{1}{16\pi G_{d+2}}\int
d^{d+1}x\sqrt{-g}\left(R-2\Lambda-\frac{1}{4}F^{2}-\frac{1}{2}m^{2}A^{2}\right),\end{equation}
with $d=3$, $\Lambda={-d(d-1)\over 2L^2}$, $m^2=d-1+(d-2)\eta^2$, where $\eta\ll1$ is used as an expansion parameter related to a small deformation of an AdS black brane,
\begin{equation}\label{metric:eta}
\begin{split}
&\hspace{1.2cm}	ds^2=-c(r)b(r)^2dt^2+{dr^2\over c(r)}+r^2(dx^2+dy^2),\\
&c=c_0(r)+\eta^2\mu^2c_1(r),~~c_0=r^2\left(1-{r_0^3\over r^3}\right),~~b=1+\eta^2\mu^2b_1(r).\\
\end{split}
\end{equation}
It is easy to see that the expansion in $\eta$ of the metric given in eq. (\ref{metric:Lifshitz:zeroT}) with $z=1+\eta^2$ may be expressed in the form of eq. (\ref{metric:eta}). The dynamical exponent is therefore $z=1+\eta^2$ and $\eta=0$ corresponds to the AdS-Schwarzchild black brane. The functions $c_1$ and $b_1$ have been given in \cite{Korovin2013b}. Moreover, \cite{Korovin2013b},
\begin{equation}
A_t=\mu\eta a_{t},~~a_t=r_0{1-u^3\over u}\Gamma\left({4\over3}\right)\Gamma\left({5\over3}\right)\pFq{2}{1}\left({1\over3},{2\over3},2,1-u^2\right),~~u={r_0\over r}
\end{equation}
where the constants in $a(u)$ are chosen so that $A_t(r)\sim\mu\eta r$ close to the boundary. 

As before, in order to study conductivity, we add perturbations on the metric and gauge field,
\begin{equation}
\delta g_{xt}=\eta^2g_{1}e^{-i\omega t},~~\delta A_x=\eta a_x e^{-i\omega t}.
\end{equation}
We note perturbations in the metric which couples to the perturbation in the gauge field enters at order $\eta^2$. The equation of the perturbation $\delta A_x$ at all orders in $\eta$ is 
\begin{equation}\label{Ax:eq:Lifeta}
A_x''+A_x'\left(-{g_{rr}\over 2g_{rr}}+{g_{xx}\over2g_{xx}}\right) +A_xg_{rr}\left(-{\omega^2\over g_{tt}}-m^2\right)+A_x{A_t'^2\over g_{tt}}=0\ ,
\end{equation}
where by $A_x$ we mean the full perturbation $\delta A_x$. Expanding the previous equation in $\eta$ we obtain to leading order the following equation for $\delta A_x$,
\begin{equation}\label{ax:eq:Lifeta}
a_x''(r)+a_x'(r){c_0'(r)\over c_0(r)}+a_x(r)\left({\omega^2\over c_0(r)^2}-{2\over c_0(r)}\right)=0\ .
\end{equation}
Clearly, the last term in eq. (\ref{Ax:eq:Lifeta}) is of ${\cal O}(\eta^3)$ and does not enter in eq. (\ref{ax:eq:Lifeta}). To obtain the Drude weight we need to solve this equation perturbatively in frequency,
\begin{equation}\label{ax:expansion}
a_x\simeq\left(1-{r_0^3\over r^3}\right)^{-i{\omega\over 3r_0}}\left(a^{(0)}_x(r)+\omega a^{(1)}_x(r)+\dots\right),
\end{equation}
and to impose on $a^{(0)}_x$ regularity at the horizon. The multiplicative term in eq. (\ref{ax:expansion}) ensures $a_x$ is purely ingoing at the horizon. The solution of $a^{(0)}_x$ is
\begin{equation}\label{Ax:Skenderis}
a^{(0)}_x=r_0^2 C \pFq{2}{1}\left(-{1\over3},{2\over3},{1\over3},{r^3\over r_0^3},\right)+r^2 \tilde C \pFq{2}{1}\left({1\over3},{4\over3},{5\over3},{r^3\over r_0^3},\right).
\end{equation}
Imposing regularity at the horizon gives,
\begin{equation}\label{c1:Skenderis}
\tilde C=-C {\Gamma\left({1\over3}\right)^2\Gamma\left({4\over3}\right)\over\Gamma\left({-1\over3}\right)\Gamma\left({2\over3}\right)\Gamma\left({5\over3}\right)}\ .
\end{equation}
We normalize $a^{(0)}_x$ by setting 
\begin{equation}\label{c2:Skenderis}
C={a_0\over  r_0^2}{2^{2/3}\sqrt{\pi}\Gamma\left({2\over3}\right)\over\Gamma\left({1\over6}\right)},
\end{equation}
so that close to the boundary $a^{(0)}_x\sim {a_0\over u}-{a_0\over3}u^2\log{u}+a_1 u^2 $. 
Moreover it has been shown in \cite{Korovin2013} that the counterterms,
\begin{equation}
S_{ct}={1\over 16\pi G}\int d^dx\sqrt{\gamma}\left(2K-{2(d-1)\over L}+{1\over2}A_\mu A^\mu\right),
\end{equation}
renormalize the boundary action up to order $\eta^2$. Using the solution, eqs. (\ref{Ax:Skenderis})-(\ref{c2:Skenderis}), we obtain a finite Drude weight at order $\eta^2$,
\begin{equation}\label{KLifshitz}
K={\alpha \over 16\pi G}{\eta^2\over36 r_0}\ ,
\end{equation}
with $\alpha = 37-\log{729}$.

In order to compare this result with the prediction of the universal Drude weight, eq. (\ref{hydroK}), we use the charge density, which may be obtained from the one-point function $\langle J_t\rangle\propto \mu \eta$ \cite{Korovin2013b}. This leads to $K_{\rm U}\propto \mu^2\eta^2$. Therefore, the prediction of the universal Drude weight is different from the direct calculation of the Drude weight which is independent of $\mu$ at ${\cal O}\left(\eta^2\right)$, eq. (\ref{KLifshitz}).

Moreover, the MS bound vanishes $K_{\rm MS} = 0$ at this order in $\eta$ since the terms coupling $\delta g_{xt}$ and $\delta A_x$ occur at ${\cal O}(\eta^3)$. In summary, non-AdS boundaries lead to a vanishing $K_{\rm MS}$ and a Drude weight different from the universal one. 

\section{Deviations from universality II: $U(1)$ symmetry breaking}\label{sec:EMDW}
We study another model in which the Drude weight is not given by the universal prediction and the MS bound is not saturated because of spontaneous symmetry breaking of the dilaton. 
We consider the following EMd theory which has been explored in detail in \cite{charmousis2010,Gouteraux2011}:
\begin{equation}\label{ActionEMD}
\begin{split}
\Semd={1\over 2\kappa^2}\int &d^{p+1}x\sqrt{-g}\left[R-{1\over 2}(\partial\phi)^2+V(\phi)-{Z(\phi)\over4}F^2\right]\ .\\
\end{split}
\end{equation}
The AdS radius is set to $L=1$ and 
\begin{equation}\label{emd_Z_V}
Z(\phi)=\ch({\gamma \phi})\ ,\ V(\phi)=-2\Lambda-{2m^2\over \delta^2}\sh^2(\delta \phi)\ ,
\end{equation}
where $\gamma,\ \delta > 0$. The UV completion of $V(\phi)$ is chosen such that no logarithmic divergences appear close to the UV, \cite{Kiritsis2015}. The UV completion of $Z(\phi)$ is fixed by requiring $Z'(\phi=0)=0$, which ensures the existence of a second order phase transition at finite temperature driven by the condensation of the dilaton. Moreover $m^2$ controls the scaling dimension of the operator dual to the dilaton in the usual way: $\Delta={1\over2}(p-\sqrt{p^2+4m^2})$. Following \cite{Kiritsis2015} we take the metric ansatz,
\begin{equation}\label{metric:ansatz}
\begin{split}
&ds^2=-D(r)dt^2+B(r)dr^2+Cd\vec{x}^2,\\
&D={g(r)\over r^2h(r)},\ \ B={1\over r^2g(r)},\ \ C={1\over r^2}\ ,\\
\end{split}
\end{equation}
where the UV is at $r=0$ and the horizon at $r_H=1$, $h(r_H)=0$. The geometry is asymptotically $AdS$, so close to the boundary,
\begin{equation}
\begin{split}
&\phi\sim{\phi_a r^\Delta}+{\phi_b r^{p-\Delta}}+\dots\ ,\\
&g\sim1+\dots+{g_p r^p}+\dots\ ,\\
&h\sim 1+\dots+{h_p r^p}+\dots\ ,\\
&A_t=\mu+{\rho  r^{p-2}}+\dots\ .\\
\end{split}
\end{equation}
We impose $\phi_b=0$, and choose $m^2=-2/L^2$, $\Delta=1$ and $p=3$. We add the usual perturbations $\delta A_x$ and $\delta g_{xt}$. In $p=3$ dimensions, the electrical conductivity is:
\begin{equation}
\sigma=\left.\sqrt{-g}{Z(\phi)\over B C}\right|_{r\to0}{A_x^{(1)}\over i\omega A_x^{(0)}}={A_x^{(1)}\over i\omega A_x^{(0)}},
\end{equation}
where $A_x\sim A_x^{(0)}+A_x^{(1)}r+\dots$.

As it was mentioned in sec. \ref{sec:Universiality}, it has recently been shown, \cite{jain2010,chakrabarti2011,Davison2015}, that in the EMd theory given by eq. (\ref{ActionEMD}), the regular part of the DC conductivity and the Drude weight may be expressed in terms of thermodynamic quantities and the electromagnetic coupling constant 
\begin{equation}\label{eq:sigmaEMD}
\sigma_{DC}^{reg}=\left({s T\over \epsilon +P}\right)^2Z_HC_H^{p-3\over2},\ \ K=K_{\rm U}={\rho^2\over\epsilon+P}\ ,
\end{equation}
where $s$ is the entropy density, $T$ temperature, $P$ pressure and $\epsilon$ energy density. The subindex $H$ indicates that the corresponding term is evaluated at the horizon and $C$ is the metric function in the general metric ansatz given in the first line of eq. (\ref{metric:ansatz}). Similarly to the Einstein-Maxwell theory, eq. (\ref{action}), the Drude weight above is given by the same expression and it saturates the MS bound.

The situation is different in the presence of a gauge field mass term in the EMd action,
\begin{equation}\label{SW}
S_W=-\int d^{p+1}x\sqrt{-g}\frac{W(\phi)}{2}A_\mu A^\mu\ ,
\end{equation}
and 
\begin{equation}\label{coupling_W}
W(\phi)=W_0\left[-1+\ch^2(\eta \phi/2)\right].
\end{equation}
We have chosen $W(\phi)$ such that $W(\phi=0)=0$ and $W'(\phi=0)=0$ to avoid divergencies in the UV.
In the following we consider the action $\Semd+S_W$ given by (\ref{ActionEMD}) and (\ref{coupling_W}).

More specifically we investigate fractionalized {\it IR-charged} solutions, \cite{Gouteraux2013,Kiritsis2015}, with a constant scalar in the IR and extremality for vanishing temperature. In the context of AdS/CFT, a fractionalized state arises when the dual field theory charge density is not determined only by the charged bulk fields but also by a horizon charged flux \cite{Hartnoll2011lectures,Hartnoll2011}.
 Recently, it has been claimed  \cite{Sachdev2010,Huijse2011} that the gravity dual of a fractionalized Fermi liquid
is a background with $AdS_2\times \mathbb{R}^n$ horizon that has a finite entropy event at zero temperature. A fractionalized state occurs for a non-vanishing electric flux in the IR, \cite{Gouteraux2013}, which in our case
\begin{equation}
\lim_{r\to r_h}{1\over4\pi}\int_{\mathbb{R}^2}Z(\phi)\ast F=\lim_{r\to r_h}-{V_{\mathbb{R}^2}\over4\pi}Z(\phi){CA_t'\over\sqrt{BD}}\neq0.
\end{equation}

\begin{figure}[H]
\center
\includegraphics[scale=0.62,clip]{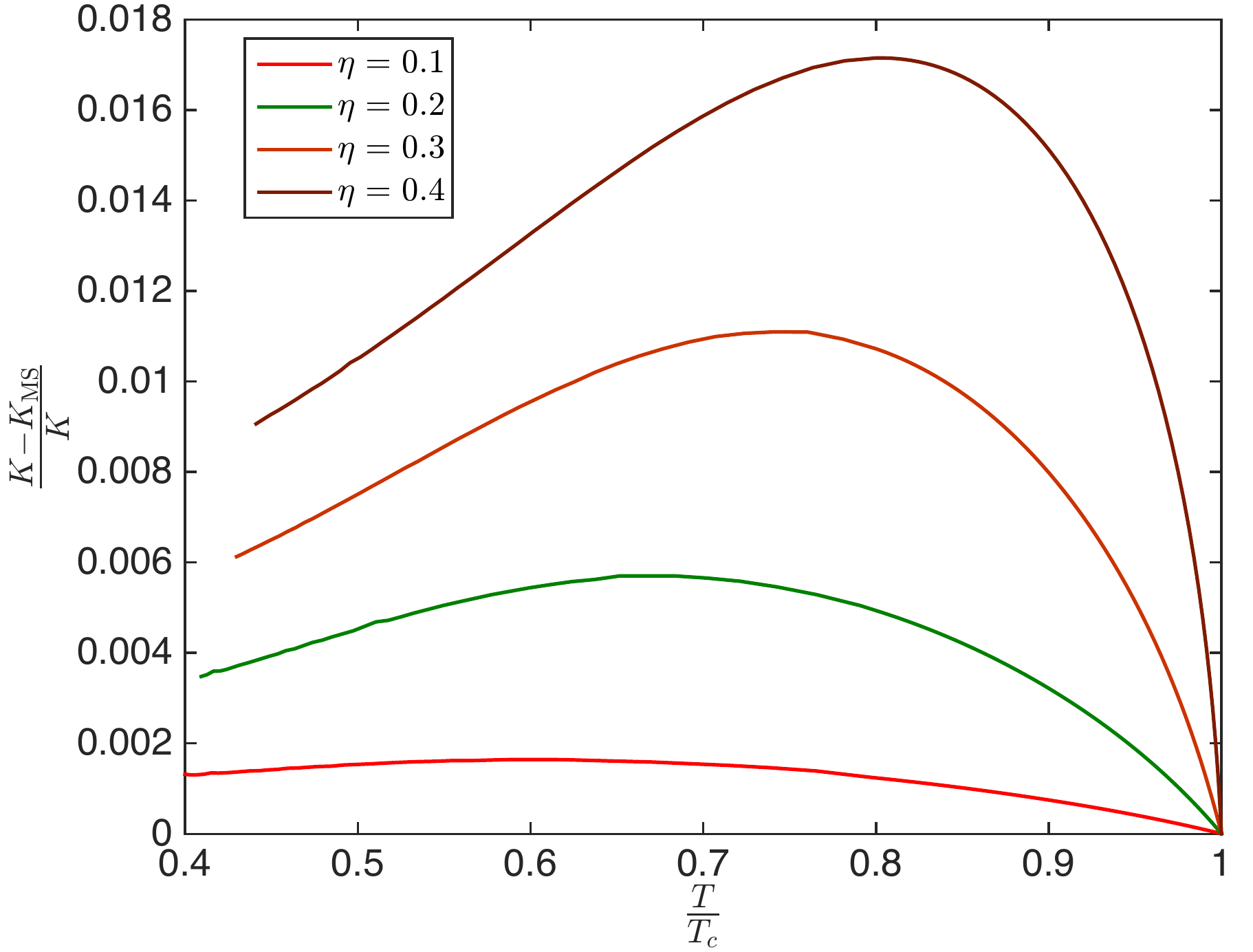}
\vspace{-0.4cm}
\caption{Difference between the Drude weight and the MS bound in the theory given by $\Semd+S_W$, eqs. (\ref{ActionEMD}) and (\ref{SW}). At the critical temperature the Drude weight is given entirely by the universal expression. This is expected since the dilaton vanishes and the background is the Reissner-Nordstr\"{o}m black hole, for which the Drude weight is given by the universal result, $K=K_{\rm MS}={\rho^2\over \epsilon+P}$. For $T<T_c$ the dilaton condensates and the physics is similar to that of holographic superconductors where the dilaton is interpreted as the modulus of a charged scalar. The Drude weight is always above the MS bound since the spontaneous breaking of the $U(1)$ symmetry produces an extra contribution proportional to the superfluid density which persists even  in the presence of momentum dissipation. The parameters used are $W_0=1$, $\gamma\delta=1$, $\delta=1/2$.  The parameter $\eta$ is given in horizon units.}
\label{KW_diff}
\end{figure}

The action $\Semd+S_W$ still has translational symmetry so we expect a finite Drude weight. Indeed 
the numerical results, depicted in Fig. \ref{KW_diff}, show the Drude weight, for $T<T_c$ where dilaton condensation occurs, is larger than the universal prediction. The MS bound, still given by eq. (\ref{suz}), is not saturated as we expect an additional contribution from the superlfluid density that does not depend on thermodynamic quantities. Similarly to holographic superconductors, \cite{Hartnoll2008}, this extra contribution is associated to the $U(1)$ spontaneous symmetry breaking, where the dilaton may be taken as the modulus of a complex scalar. With respect to the transport properties, the main difference\footnote{The potential of the scalar field is not quadratic for the EMd model and the gauge field coupling is not constant.} with respect to holographic superconductor narrows down to the different coupling between the gauge field and the dilaton. While in our model it is given by (\ref{coupling_W}), for holographic superconductors it is quadratic in the scalar field with a coupling strength proportional to its charge.

\begin{figure}[H]
\center
\includegraphics[scale=0.62,clip]{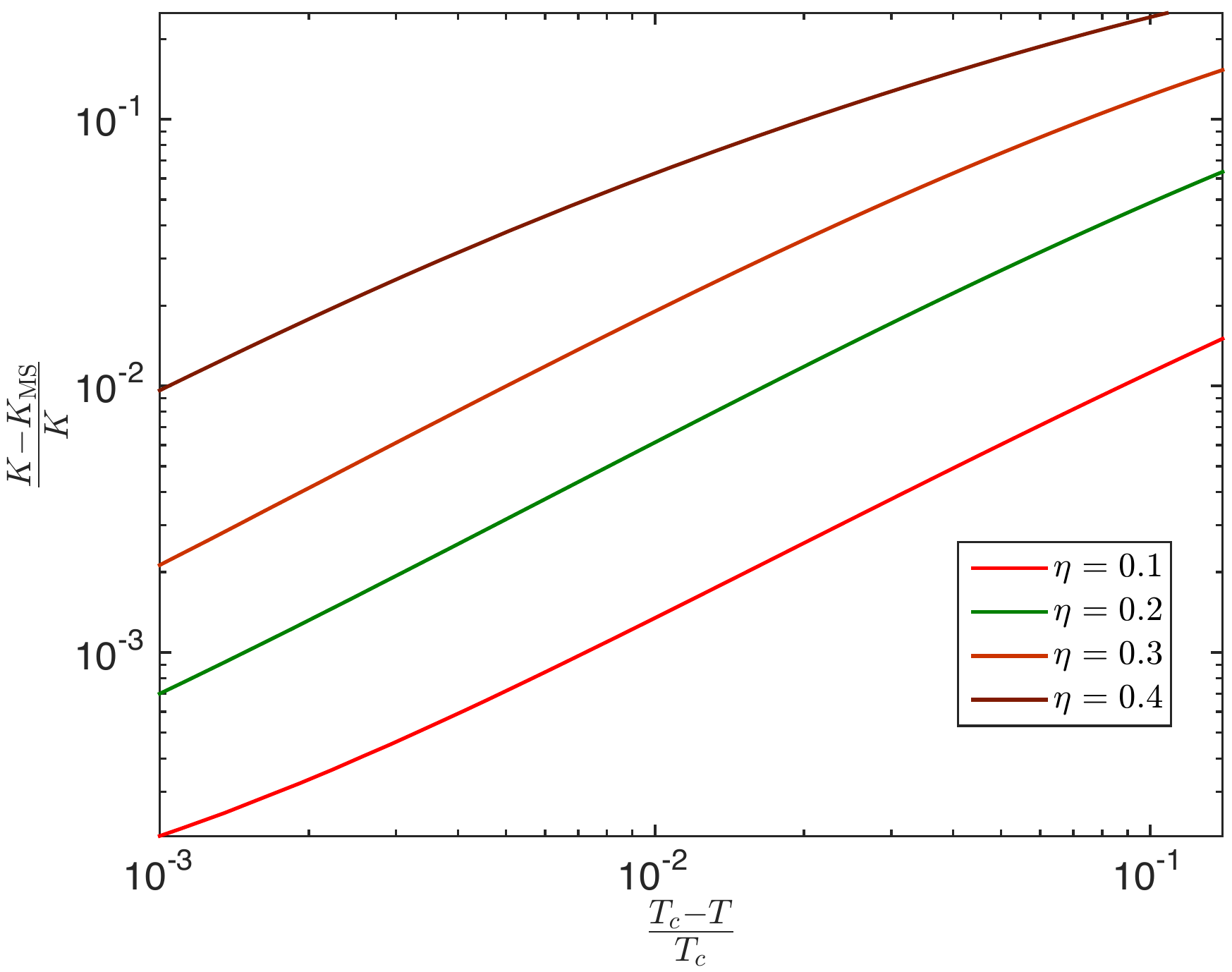}
\vspace{-0.4cm}
\caption{Difference between the Drude weight and the MS bound in the theory given by $\Semd+S_W$, see eqs. (\ref{ActionEMD}) and (\ref{SW}). Close to the critical temperature the mass term coupling $W(\phi)\sim \eta^2\phi^2+\dots$, in which case the model is similar to that of standard holographic superconductors. Therefore, the extra contribution to the Drude is expected to be proportional to $\langle O_1\rangle^2\sim {T-T_c\over T_c}$. Consequently, for some range of temperatures, the slopes of the lines shown in logarithmic scale are similar. For larger values of $\eta\phi$ (either far from $T_c$ or for larger $\eta$) the mass coupling receives higher order corrections, which affect the extra contribution to the Drude weight. Thus, for a fixed $\eta$, deviations from a linear behaviour are observed by increasing $T-T_c\over T_c$  (increasing expectation value of the dilaton), . Similarly, for larger $\eta$, the linear behaviour occurs closer and closer to $T_c$. The parameters used are the same as in Fig. (\ref{KW_diff}).}
\label{KW_O1}
\end{figure}

Moreover, at least close to the transition temperature, it is expected the Drude weight to be determined by two additive contributions. The universal one, given by $\rho^2\over \epsilon+P$, and another one proportional to the superfluid density $n_s \propto \langle O_1\rangle^2$ where $\langle O_1\rangle$ is the expectation value of the operator dual to the dilaton. We also expect that the transition is controlled by mean field critical exponents, $\langle O_1\rangle\propto (T-T_c)^{1/2}$. The results  of Fig. \ref{KW_O1}, confirm these predictions: close to $T_c$ the extra contribution to the Drude weight is linear in $T-T_c\over T_c$. We use logarithmic scale since the region of temperatures where the linear scaling is observed is small.

\section{Momentum dissipation, scattering time and bounds on the conductivity}\label{sec:axions}
In this section we study the DC conductivity in systems where translational invariance is weakly broken. If the breaking is sufficiently weak, so that the scattering time $\tau$ is sufficiently long, we still expect the Drude weight $K$, or more precisely the part of it related to conservation of momentum, of the translationally invariant theory to still control the DC conductivity, 
 \begin{equation}\label{tauk}
 \sigmadc \approx K \tau.
\end{equation}   

We confirm the validity of (\ref{tauk}) by computing explicitly the scattering time $\tau$ which is nothing but the \textit{dominant} pole of the relevant Green's function that controls the decay of momentum. The poles of the Green's functions are obtained from the quasinormal modes of the corresponding metric and field perturbations, \cite{Kaminski2010}. The \textit{dominant} pole of the Green's functions corresponds to the purely imaginary pole with the smallest imaginary part that describes the slowest decaying mode of the system. This is the only one which is relevant in the limit of weak momentum dissipation.

We employ the following Einstein-Maxwell-axion action, \cite{Bardoux2012,Andrade2014}, to model momentum dissipation,
\begin{equation}\label{the model}
\begin{split}
	S_0 &= \int_M \sqrt{-g} \left[ R - 2 \Lambda - \frac{1}{2} \sum_{I}^{d-1} (\partial \psi_I)^2  - \frac{1}{4} F^2 \right ] d^{d+1} x,\\ 
\end{split}
\end{equation}
where for convenience we have omitted the counterterms needed to regularize the action in the boundary.
In order to proceed we turn on a perturbation of the gauge field, $\delta A_x =e^{-i\omega t}a_x(r)$. For the axion model this perturbation couples to a metric and a scalar perturbation $\delta g_{xt}=e^{-i\omega t}r^2H_{tx}(r)$, $\delta \psi=e^{-i\omega t}\alpha^{-1}\chi(r)$, \cite{Andrade2014}, where $r\in(r_0,\infty)$ is the holographic radial coordinate and $\alpha$ is the parameter related to the breaking of translational symmetry. The equations for these perturbations at zero spatial momentum are given in Ref. \cite{Andrade2014} for arbitrary bulk dimensions $d+1$:
\begin{align}
\label{eq1:qnm:RN}
&a_x'' + \left[ {f'\over f} + {(d-3)\over r} \right] a_x' + {\omega^2\over f^2} a_x  + {\mu (d-2)\over f} {r_0^{d-2}\over r^{d-3}} H_{tx}' = 0\ , \\
\label{eq2:qnm:RN}
&\chi'' + \left[ {f'\over f} + {(d-1)\over r} \right] \chi' + {\omega^2\over f^2} \chi - {i \omega \alpha^2\over f^2} H_{tx} = 0\ , \\
\label{eq3:qnm:RN}
&{i \omega r^2\over f} H_{tx}' + {i \omega \mu (d-2) \over f} {r_0^{d-2}\over r^{d-1}} a_x  -  \chi' = 0\ .
\end{align}

In general, the dominant quasinormal mode, and therefore $\tau$ can only be computed numerically. However, in the limit $\alpha \ll T$ an analytical expression for $\tau$, associated to the transverse fluctuations above, was found for $d=3$ \cite{Davison2015}, 
\begin{equation}\label{tauAxion}
\tau^{-1}\simeq\eta {\alpha^2\over \epsilon +P}={\alpha^2 \over 3r_0\left[1+{\mu^2\over 4r_0^2}\right]}\ .
\end{equation}
We note that this expression is identical to that obtained in the context of massive gravity \cite{Andrade2014,Davison2013} with the replacement $\alpha^2 \to 2m^2$  where $m$ is the mass of the graviton. It is still unclear to what extent this dependence on temperature is shared by other models. 
By following the approach of \cite{Davison2015} we have generalized this expression to $d>3$,
\begin{equation}\label{tauAxiond}
\tau^{-1}\simeq\eta {\alpha^2\over \epsilon +P}={\alpha^2 \over r_0d\left[1+{(d-2)\mu^2\over 2(d-1)r_0^2}\right]}\ .
\end{equation}
This expression, is valid only for weakly breaking of translational, namely, up to ${\cal O}(\alpha^4/T^4)$ corrections. 
An obvious correction ${\cal O}(\alpha^4)$ is obtained by substituting the energy density and pressure corresponding to the system with $\alpha\neq0$, however, as shown in \cite{Davison2015}, this is not the only one. We are not interested in such corrections and refer to \cite{Davison2015} for details.

We also evaluate $\tau$ numerically following the method proposed in \cite{Kaminski2010}, which consists in using independent sets of boundary conditions ${\mbox{BC}_i},\ i=1,\dots,N$, in the IR to find various solutions in the UV. One constructs the matrix which has in each column the boundary value of the fields $\{\phi^{\mbox{BC}_k}_n\},\ n=1,\dots,N$, with a given set of IR boundary condition, ${\mbox{BC}_k}$, and in every row a field with each boundary condition $\phi^{\mbox{BC}_i}_n$, $i=1,\dots,N$, i.e., for $N$ fields the matrix is $N\times N$. The leading quasinormal mode $\tau$ is given by the purely imaginary frequency, with smallest absolute value, for which the determinant of such matrix vanishes.

\begin{figure}[H]
\center
\includegraphics[scale=0.55,clip]{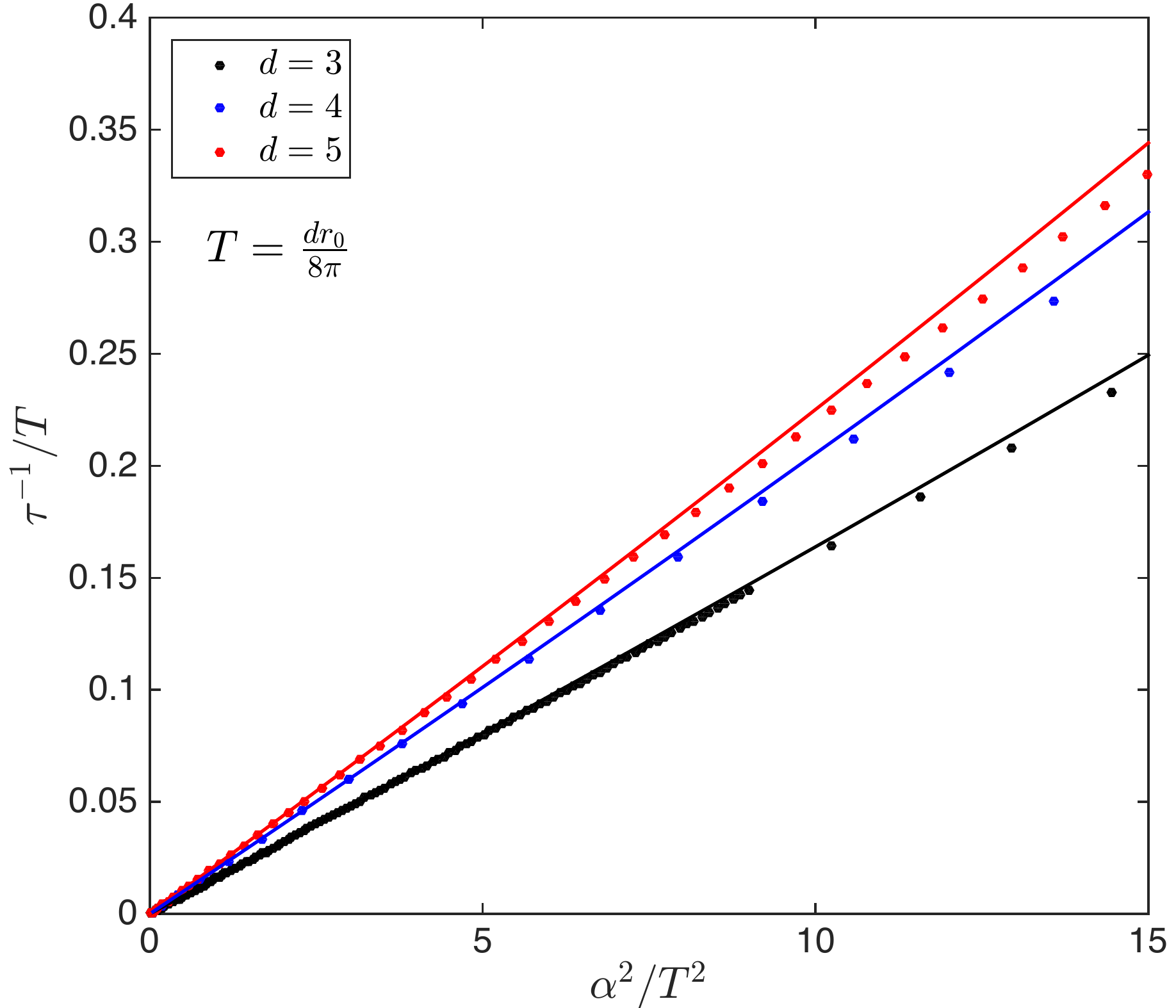}
\vspace{-0.4cm}
\caption{Scattering time in the Einstein-Maxwell-axion model, eq. (\ref{the model}), at fixed temperature in different number of dimensions. The dots are numerical results while the lines are obtained from the analytical expression proposed, eq. (\ref{tauAxiond}), valid until second order in $\alpha$. As anticipated in \cite{Davison2015b}, the analytical expression deviates from the numerical results progressively as the contribution of additional quasinormal modes to the total scattering rate increases continuously for larger $\alpha/T$.}
\label{tau_d345}
\end{figure}

The numerical results, depicted in Fig. \ref{tau_d345}, are very close to the analytical prediction (\ref{tauAxiond}) even beyond  its limit of applicability, $\alpha \ll T$. 
Interestingly, the dependence of $\tau$ on dimensionality is rather weak. In the high temperature limit, assuming  $\mu/T \ll 1$, $\tau \propto T$ for all $d$'s. We find hard to interpret physically this linear dependence on temperature.  The temperature dependence of the scattering time is very sensitive to the source of scattering (phonons, impurities, electrons), the range of temperatures and whether the material is an insulator, metal or semiconductor. 
Sometimes, it increases with temperature, as for charge impurities in semiconductors. In many other cases decrease with temperature as for phonon scattering at high temperature. However, we are not aware of any simple situation in which is linear.  It would be interesting to find a holographic model in which the scattering time has a richer temperature dependence.

We have now all the information to compute the DC conductivity. For sufficiently small $\alpha/T$, from eqs. (\ref{tauAxiond}) and (\ref{K_hydro}) with $z_0=1/r_0$, $\sigma_{DC} \approx K \tau \approx \mu^2(d-2)^2r_0^{d-3}/\alpha^2$. Not surprisingly, except for the incoherent contribution which is smal in this limit, this is the analytical result already obtained in \citep{Andrade2014}. 

Since $K$ is constrained by the MS bound, the conductivity, for a fixed large $\tau$, has also a lower bound $\sigma_{\rm DC} \geq K_{\rm MS}\tau$. However, the bound is trivial here because the MS bound is saturated in this model. 
We shall see a different behavior in the next section when we study Einstein-Maxwell-dilaton actions.

\subsection{Momentum relaxation and scattering time $\tau$ in Einstein-Maxwell-dilaton backgrounds}\label{sec:emd:dilaton}
\begin{figure}[H]
\center
\includegraphics[scale=0.62,clip]{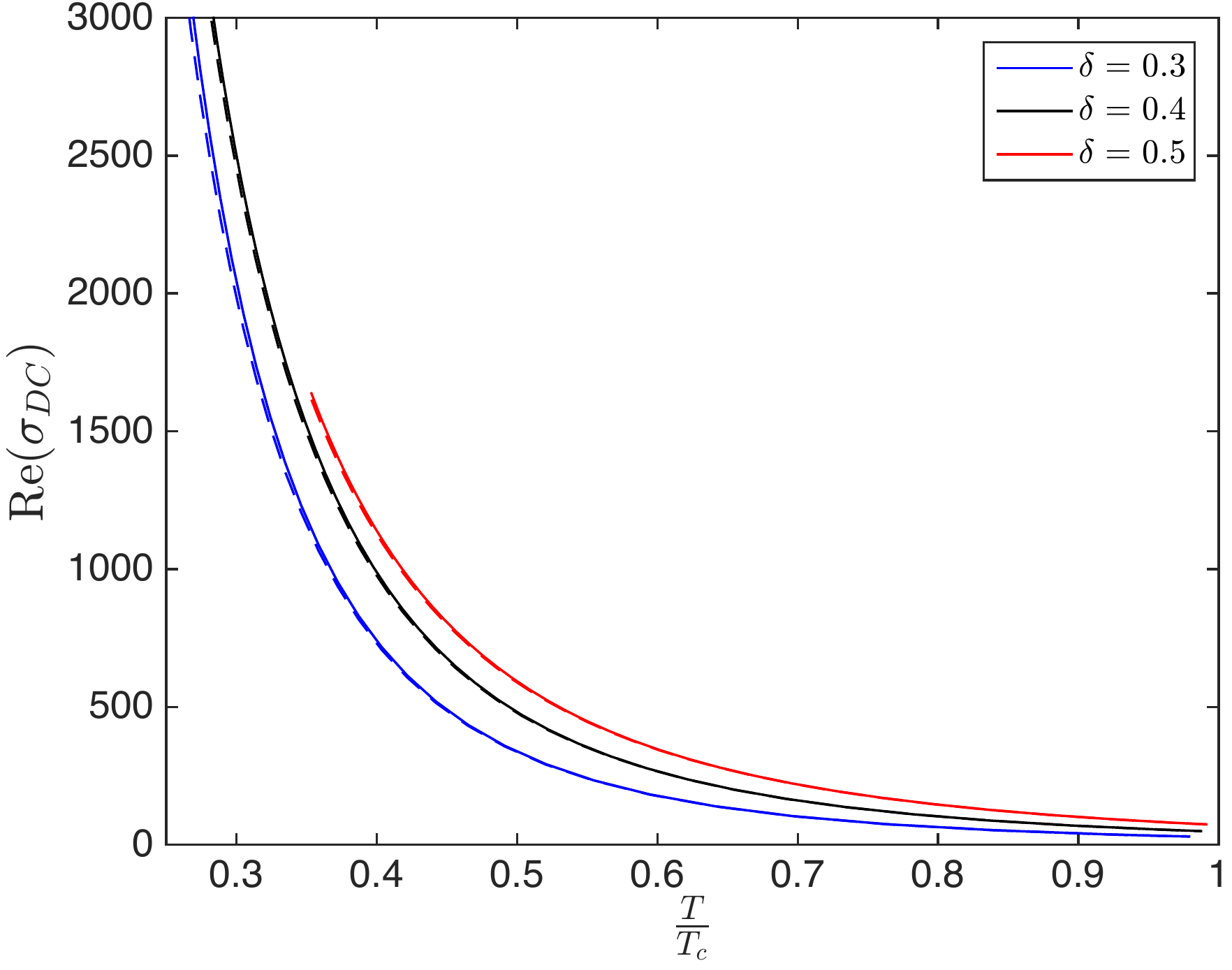}
\vspace{-0.4cm}
\caption{Real part of the DC conductivity the theory $S_{\text{EMd}}+\Sax$. The continuous lines correspond to the numerical result. The dashed lines are obtained from eq. (\ref{sigma_emd}). Here, we have chosen $\gamma\delta=1$, $\lambda=1/2$, $\alpha/r_0=0.1$. Clearly, for weak breaking of translational invariance ($\alpha\ll T$), the conductivity has a contribution controlled by the MS bound of the translationally invariant theory ($\Semd$). As for the Einstein-Maxwell theory, the MS bound is saturated by the universal result, $K_{\rm MS}={\rho^2\over\epsilon+P}$.}
\label{DC_Kemd}
\end{figure}

We now repeat this analysis in a more general EMd theory $\Semd+\Sax$, 
\begin{equation}\label{SY}
\begin{split}
S_{\text{axion}}=&-{1\over 2\kappa^2}\int d^{p+1}x\sqrt{-g}\frac{Y(\phi)}{2}\sum_{i=1}^{p-1}(\partial\psi_i)^2\ ,\\
\end{split}
\end{equation}
with,
\begin{equation}\label{coupling_Y}
Y(\phi)=-1+2\ch^2(\lambda \phi),\quad \psi_i=\alpha x_i,\quad i=1,\dots,p-1\ .
\end{equation}
In this theory the conductivity at zero frequency is finite. It is obtained analytically by finding the massless mode of the system of equations for the perturbations of the gauge field, metric and axions, $\delta a_x$, $\delta g_{xt}$ and $\delta \psi_x$. This allows to decouple the system of equations and compute $\sigma_{DC}$ analytically. This approach was first introduced in \cite{Blake2013} for a model of massive gravity and later applied to $\Semd+S_{\rm axion}$ in, \cite{Gouteraux2014,Kiritsis2015}. We simply cite the final result,
\begin{equation}\label{gout}
\sigmadc=Z_HC_H^{p-3\over2}+{\rho^2\over \alpha^2 C_H^{p-1\over2}Y_H}\ ,
\end{equation}
 where $\rho$ is the charge density, $\alpha$ is defined in eq. (\ref{coupling_Y}) and, again, the subindex $H$ means the corresponding quantity is evaluated at the horizon. As in the previous section we now compare this result for $p=3$ with
\begin{equation}\label{sigma_emd}
\sigmadc=Z_H+K_{\rm MS}\tau\ ,
\end{equation}
which, as we will see, is easier to interpret physically. $K_{\rm MS}$, the MS bound, is calculated from  eq. (\ref{suz}) with a single conserved quantity associated to momentum conservation in the theory with axions turned off, $\Semd$, at the same temperature and charge density. It coincides with the universal value $K_{\rm MS}=K_{\rm U}$, eq. (\ref{Kuniv}). The scattering time, $\tau$, is again computed from the dominant quasinormal model of the theory $\Semd+\Sax$ as explained in the previous sections. More specifically, we add perturbations of the gauge field, metric and axion in the theory given by $\Semd+\Sax$ and solve for the dominant quasinormal mode using the the  equations analogous to those given in eqs. (\ref{eq1:qnm:RN})-(\ref{eq3:qnm:RN}) for the Einstein-Maxwell theory. We follow the same method explained in sec. \ref{sec:axions} to compute the dominant quasinormal mode. 
The results, depicted in Fig. (\ref{DC_Kemd}), clearly show that again in this case eq. (\ref{sigma_emd}) provides an excellent description of the DC-conductivity eq. (\ref{gout}) in the $\Semd+S_{\text{axion}}$ model.

As in the previous case the MS bound is saturated and therefore the associated bound of the conductivity $\text{Re}(\sigma_{\rm DC}) \geq Z_HC_H^{p-3\over2}+K_{\rm MS}\tau$ is not of special relevance.  In light of these results, it is not difficult to understand that, once the axions are switched on and translational invariance is weakly broken, the Drude weight of the translational invariant theory still controls the coherent part of the DC conductivity.
Similar results hold for theories $\Semd+\Sax$ with $Z'(\phi=0)\neq0$, which corresponds to a black hole with dilaton condensation for all temperatures.

\subsection{Drude weight and momentum relaxation in theories with $U(1)$ symmetry breaking}
\begin{figure}[H]
\center
\includegraphics[scale=0.62,clip]{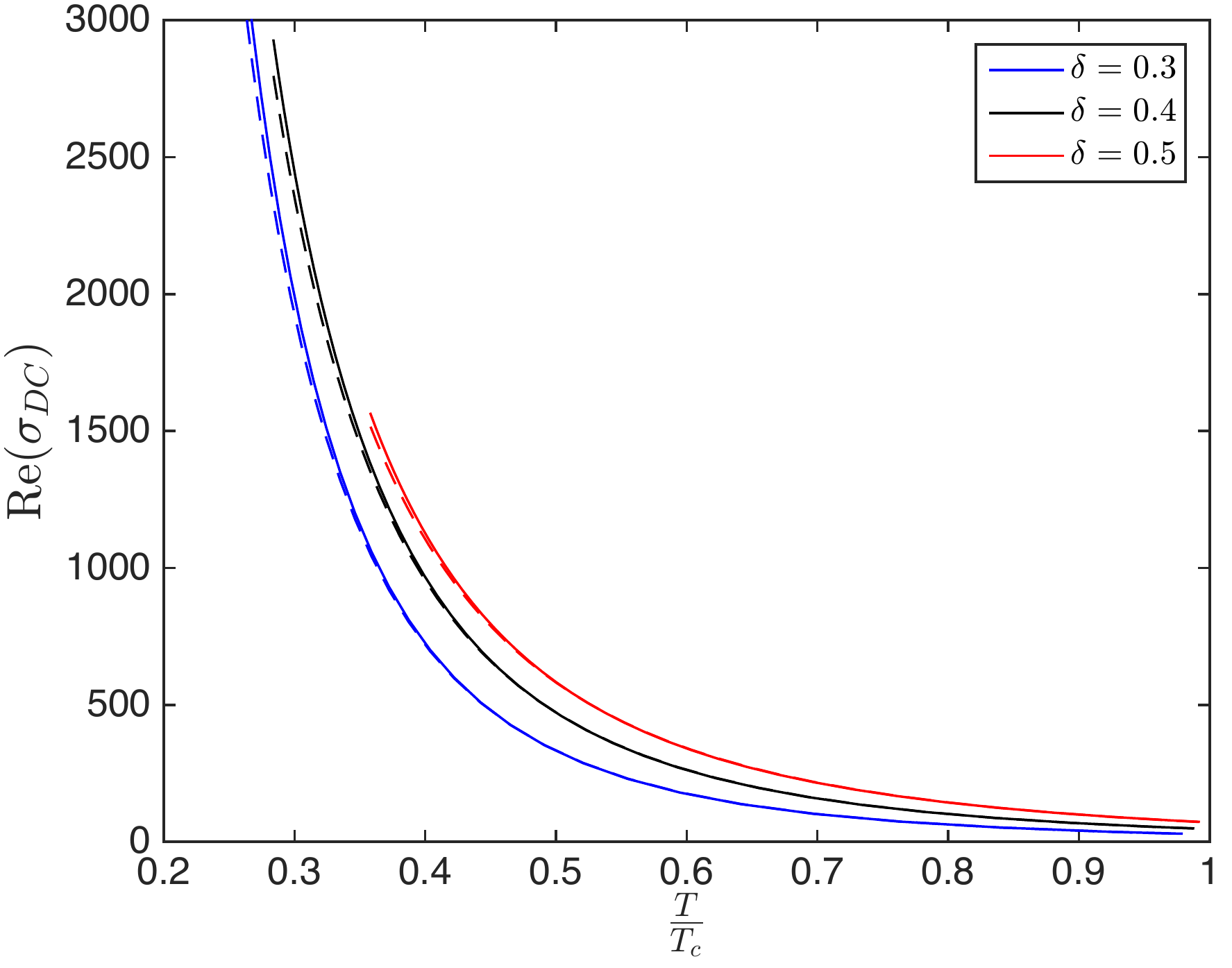}
\vspace{-0.4cm}
\caption{Regular part of the DC conductivity in the theory given by $\Semd+S_W+\Sax$, see eqs. (\ref{ActionEMD}), (\ref{SW}) and (\ref{SY}). The continuous lines are numerical results while the dashed lines correspond to eq. (\ref{sigma_EMDWY}). Clearly, the regular part of DC conductivity is controlled by the MS bound of the theory $\Semd+S_W$, while the superfluid mode still contributes to a finite Drude weight even when translational symmetry is broken. The parameters used are $\gamma\delta=1$, $\delta=1/2$, $W_0=1$, $\eta/r_0=0.3$, $\lambda=1/2$, $\alpha/r_0=0.1$.}
\label{DC_EMDWY}
\end{figure}

In this section, we study the DC conductivity in the following theory with spontaneous $U(1)$ symmetry breaking and weak momentum dissipation,
\begin{equation}\label{SEMDWY}
S=\Semd+S_W+\Sax\ ,
\end{equation}
together with eqs. (\ref{ActionEMD}) (\ref{SW}) and (\ref{SY}) and the couplings given in eqs. (\ref{emd_Z_V}), (\ref{coupling_W}) and (\ref{coupling_Y}).
As discussed in secs. (\ref{sec:Universiality}) and (\ref{sec:EMDW}) we have seen that in the absence of axions, $\Semd+S_W$, the Drude weight receives an extra contribution from the superfluid mode and $K > K_{\rm MS}$. In Fig. (\ref{DC_EMDWY}) we show that in the presence of axions, which break translational symmetry, the DC-conductivity of the dual theory to the gravity action eq. (\ref{SEMDWY}) is controlled by the MS bound, $K_{\rm MS}$, instead of by the Drude weight $K$ of the theory in the absence of axions, $\Semd+S_W$. In other words, 
\begin{equation}\label{sigma_EMDWY}
{\rm Re}\left(\sigma_{\rm DC}^{\rm reg}\right)=Z_H+K_{\rm MS}\tau\ ,
\end{equation}
where again $K_{\rm MS}=K_{\rm U}$ and the scattering time, $\tau$, is computed from the dominant quasinormal model of the theory eq. (\ref{SEMDWY}). 

The results depicted Fig. \ref{DC_EMDWY} show that, similarly to the EMd model+axion studied in sec. \ref{sec:emd:dilaton}, eq. (\ref{sigma_EMDWY}) indeed describes the DC-conductivity. In the theory given by eq. (\ref{SEMDWY}), despite the fact momentum is not conserved, the Drude weight is not zero because the superfluid density is finite for sufficiently low temperatures. Therefore, only the part of the Drude weight that disappears once the axions are switched on, contributes to the DC conductivity. The bound on the DC conductivity associated to the MS bound is more relevant in this case as only a part of the Drude weight, the MS bound, contributes to the conductivity.

\section{Conclusions}
We have studied the Drude weight and the associated MS bound in a broad range of holographic theories. We have extended the universality of the Drude weight to the case of several massless gauge fields. We have shown that the MS bound is saturated only if the Drude weight is given by the universal expression first obtained in \cite{jain2009,Davison2015a}. For non-relativistic theories the Drude weight is finite, but different from the universal one, and the MS bound vanishes. In theories with spontaneous $U(1)$ symmetry breaking the Drude weight is larger than the universal prediction and the MS bound is finite but it is not saturated. Finally, in the limit of weak breaking of momentum conservation we have shown that the coherent part of the DC conductivity in EMd-axion theories is controlled by the leading quasinormal mode and the MS bound which suggests a lower bound, depending on the scattering time, for the conductivity as well.  
\acknowledgments{A. R. B. thanks Kostas Skenderis and Yegor Korovin for illuminating discussions. A. M. G. thanks Carlos Hoyos for illuminating discussions and acknowledges partial support from
EPSRC, grant No. EP/I004637/1. A. R. B has been supported by the Department of Physics of the University of Cambridge. Both authors are grateful to the Galileo Galilei Institute for Theoretical Physics for the hospitality and the INFN for partial support during the completion of this work.
}

\bibliographystyle{plain}
\bibliography{Mazur_bib}

\end{document}